\documentclass[final,11pt]{amsart}
\pdfoutput=1

\usepackage{amssymb}
\usepackage{amsmath}
\usepackage{lineno}
\usepackage{xcolor,mydef,url}
\usepackage{graphicx,enumerate}

\begin{document}



\title[Biofilm growth: experiments, computations, and upscaling]
{Biofilm growth in porous media: experiments, computational
  modeling at the porescale, and upscaling}


\author[M.~Peszynska]{Malgorzata Peszynska}
\address{M.~Peszynska,\\ Mathematics, Oregon State University, USA}
\email{mpesz@math.oregonstate.edu}
\author[A.~Trykozko]{Anna Trykozko}
\address{A.~Trykozko, \\ Interdisciplinary Centre for Modeling, University of Warsaw, Poland}
\email{A.Trykozko@icm.edu.pl}
\author[G.~Iltis]{Gabriel Iltis}
\address{G. Iltis, \\Brookhaven National Laboratory, USA}
\email{giltis@bnl.gov}
\author[S.~Schlueter]{Steffen Schlueter}
\address{S.~Schlueter, \\Helmholtz Centre
  for Environmental Research, Germany}
\email{steffen.schlueter@ufz.de}
\author[D.~Wildenschild]{Dorthe Wildenschild}
\address{D. Wildenschild, \\Chemical, Biological, and Environmental Engineering,
  Oregon State University, USA}
\email{dorthe.wildenschild@oregonstate.edu}

\begin{abstract}
Biofilm growth changes many physical properties of porous media such
as porosity, permeability and mass transport parameters. The growth
depends on various environmental conditions, and in particular, on
flow rates. Modeling the evolution of such properties is difficult
both at the porescale where the phase morphology can be distinguished,
as well as during upscaling to the corescale effective
properties. Experimental data on biofilm growth is also limited
because its collection can interfere with the growth, while imaging
itself presents challenges. 

In this paper we combine insight from imaging, experiments, and
numerical simulations and visualization. The experimental dataset is
based on glass beads domain inoculated by biomass which is subjected
to various flow conditions promoting the growth of biomass and the
appearance of a biofilm phase. The domain is imaged and the imaging data
is used directly by a computational model for flow and transport.  The
results of the computational flow model are upscaled to produce
conductivities which compare well with the experimentally obtained
hydraulic properties of the medium. The flow model is also coupled to
a newly developed biomass--nutrient growth model, and the model
reproduces morphologies qualitatively similar to those observed in the
experiment.
\hrule
\medskip
{\copyright}2015. This manuscript version is made available under the CC-BY-NC-ND 4.0 license http://creativecommons.org/licenses/by-nc-nd/4.0/
\\ 
Advances in Water Resources (2015), \\ http://dx.doi.org/10.1016/j.advwatres.2015.07.008.\end{abstract}

\maketitle


\section{Introduction}
\label{sec:intro}

Biofilm growth changes many physical properties of porous media such
as porosity, permeability and mass transport parameters. The goal of
the experiment and computations discussed here was to understand how
microbial species grow at different flow rates, and how this affects
the flow properties at porescale and at Darcy scale.  Numerous
studies with similar goals were reported before, see, e.g.,
\cite{Baveye98,Seifert}, but not with full 3D porescale column imaging
combined with computational modeling of hydrodynamics undertaken
here. Our paper is a step towards a fully coupled dynamic pore-to-core
scale model in which the local dynamics of flow and transport
including biomass growth is accounted for, and the simulations are
based on, and calibrated with, the experimental data.

The experiment alone cannot provide the imaging data for the biofilm
dynamics at intermediate time steps, since the amount of radiation
that the organisms are exposed to during imaging will either kill or
severely damage DNA and leads to incorrect estimates of growth
patterns. In lieu of the imaging or experiment, one can set up
simulations, and their resolution and complexity can be adapted to the
needs of a particular study. In turn, the computations are very
sensitive to the parameters chosen and to the modeling assumptions,
and these can give useless results in unrealistic geometries or with
ad-hoc parameters. Therefore, the computational model should be
fine-tuned using experimental data.

Imaging of biofilm presents its own challenges. In
\cite{ISWW,Davit,Iltis13} we described the process of imaging biofilm
growth at porescale using x-ray microtomography, a technique well
suited to three-dimensional imaging of opaque porous media; see
\cite{WilShe13} for overview of the techniques. The primary
difficulties associated with imaging include differentiation of
biofilm from the aqueous phase, both of which have similar composition
and density and therefore similar x--ray absorption
characteristics. In addition, conventional x-ray contrast agents
(e.g., potassium iodide) diffuse readily into biofilm. These issues
were resolved by using an insoluble barium sulfate particle suspension
injected directly into the hydraulically available pore space (flowing
phase) domain.

The experiment discussed here was performed at several selected flow
rates from low (linear laminar) to high (nonlinear laminar), and we
evaluated the different biofilm morphologies associated with different
flow rates, where the nutrient concentrations in the influent were
constant. The imaging at the end of the experiment provided the
geometries of the glass beads (rock) domain, the biofilm domain, and
the flowing phase domain. The imaging data was used to create grids
for the porescale simulations. With these grids we set up
computational models for the flow and transport, and upscaling.

First, we consider the flow itself. The hydrodynamics flow model
denoted below by (H) is combined with upscaling following our prior
work in \cite{PTA09,PTS10,PT13,TP13}, and is applied to the porescale
geometries with and without biofilm. We show substantial anisotropic
decrease of conductivities due to clogging and with increasing flow
rates, and we compare the computationally obtained values with those
known from the experiment. Our results are comparable to the very
recent results on larger columns in \cite{Scheibe15}, but are unique
as concerns the fine voxel resolution.

Next, we compare the biofilm morphologies observed in the experiment
with those simulated by our newly developed biomass-nutrient (BN)
model.  The (BN) model accounts for multiple phases and species: (1)
surface attached biofilm (EPS), (2) the flowing phase, and (3)
planktonic biomass transported within the flowing phase. While other
complex models have been formulated
\cite{ES,EPL,ZKlapper10,coganI,coganII,TangValocchi}, they have
features that make them difficult to apply in realistic pore
geometries. In particular, they do not include mass transport other
than that due to biofilm spreading, and/or have degenerate and
singular behavior as well as explicit treatment of nonlinearities. In
contrast, our (BN) model is fairly easy to implement and robust, yet
can account for multiple phases and species and their transport. It is
also amenable to rigorous mathematical and numerical analysis; see the
forthcoming paper \cite{PTbio}. Furthermore, the coupled (H-BN) model
used in this paper treats hydrodynamics (H) and the biofilm/nutrient
dynamics model (BN) in a time-staggered fashion. This approach allows
the local biofilm geometry, thus the fluid domain, to change.

The outline of the paper is as follows. In Sec.~\ref{sec:exp} we
overview the experiments and the data from imaging. In
Sec.~\ref{sec:flow} we describe the porescale flow simulations and
compare their results to the experiment. In Sec.~\ref{sec:bio} we
define the biofilm growth model and its numerical implementation. In
Sec.~\ref{sec:simulations} we report the numerical simulation results
of the flow and transport (H-BN) model, and discuss them in view of
the experimental results. Sec.~\ref{sec:conclusions} consists of a
summary and conclusions.

We use the following notation and nomenclature throughout the paper.
We use rectangular grid cells in 2D, and regular hexahedral or cuboids
grid cells in 3D, also referred to as rectangles. When defining
domains of flow and transport we use the notation $A\setdef{B}$ to
define the set $A$ to be the interior of the closure of the set $B$;
this allows us to include interfaces between some disjoint but
adjacent open sets in their union. When referring to the sets, we
denote by $\abs{A}$ the number of voxels or computational grid cells
covering the set $A$, and by $\chi_A$ its characteristics function
equal to one in the set and zero outside. If $q$ is some quantity, we
denote by $q^*$ its value measured in the experiment. By $q_0$ (or
$q^*_0$) we refer to its value at the initial time $t=0$, and by $q_T$ (or
$q^*_T$) we denote the values corresponding to $t=T$, the end of
experiment or simulations.

\section{Flow: experiments, imaging, simulations, and upscaling} 
\label{sec:exp}
 
In this section we discuss the flow and imaging experiments and the
process of obtaining data for the flow and transport simulations.  

\subsection{Experimental set-up and notation}

\begin{table}
\begin{center}
\begin{tabular}{|l|c|c|l|l}
\hline
Columns&Flow rate $Q$&Re&$v_{in}$[m/s]\\
\hline
Column 1&450 mL/h &Re = 10&$10^{-2}$\\
Column 8&45 mL/h &Re = 1&$10^{-3}$\\
Column 7&4.5 mL/h &Re = 0.1&$10^{-4}$\\
\hline
\end{tabular}
\end{center}
\caption{\label{tab:flow} The columns used in the experiment with the
  corresponding flow rates. The right column gives the inlet
  velocities used in the simulations which correspond (within 3\%) to
  the flow rates. The Reynolds number is computed with the formula Re$
  = \frac {\rho {q} d} {\mu {A \phi}}$, where $q$ is the flow rate and
  $A$ cross-sectional area, $\phi$ is the porosity, and $d$ is the
  average grain size. The numbering of columns follows that in
  \cite{ISWW}.}
\end{table}

The column reactors used in the experiment were 6.3mm in diameter, 30mm
long, and they were filled with soda-lime silica glass beads of size
1.4--1.7{mm}, with specific gravity 2.5. There were six columns in the
experiment reported in \cite{ISWW}. Here we report only on one column
per each of the three flow rates, denoted as $\Omega^{c}$ where
$c=1,8,7$ runs over columns; see Tab.~\ref{tab:flow}. The average
initial porosity of columns was
\ba
\label{eq:phi0}
\phi_0^*&=&39.5\%;
\ea
the hydraulic conductivity will be discussed in Sec.~\ref{sec:darcy}.

The microbial species \emph{Shewanella oneidensis MR-1}, a metal
reducing bacterium, was used to inoculate the columns \cite{Iltis13}. Over a period
of time $T=11~\mathrm{days}$, the microbes were provided nutrient
(tryptic soy broth=TSB) as well as dissolved oxygen (DO) to promote
biomass growth under various flow rates. During the experiment the
biomass formed the biofilm phase, which clogged the porespace and
changed the flow patterns. At $t=T$, the growth was stopped, and the
columns were imaged at the Advanced Photon Source, Beam-line 13-BMD
(GSECARS). Images were collected at the end of the growth period, and
three-dimensional volumes of greyscale data was processed.

The density of biofilm, while only slightly higher than that of water,
presents a challenge to imaging; see \cite{Iltis13,Davit}. The
technique described in \cite{ISWW} was to use a contrasting agent,
barium sulphate, which is physically excluded from the biofilm domain
as well as from the interior of glass beads. The principle of physical
exclusion enables the imaging, and is the basis for the interpretation of
the biofilm phase in our flow and transport model and simulations
described in Sec.~\ref{sec:flow}--\ref{sec:bio}.

In what follows we denote the space occupied by the glass beads within
the column as $\Omega_r$ (``rock''). The space occupied by biofilm is
denoted by $\Omega_b$, and that by the fluid flowing outside the
biofilm by $\Omega_f$,
\ba 
\label{eq:barium}
\Omega_f \setdef {\mathrm{\ the\ domain\ where\ barium\ agent\ is\ visible}},
\ea
and we have
$
\Omega_b \setdef {\Omega \setminus \Omega_r \setminus \Omega_f}.
$
Due to biofilm growth, both of $\Omega_f, \Omega_b$ change with time, while $\Omega_r$
does not. 

In the process of imaging, each column $\Omega$ is embedded in the
union of $610 \times 610 \times 2833$ voxels, each of volume
$(h_{\omega})^3$, with voxel size $h_{\omega}=10.5\cdot 10^{-6}$~m.
We can write that $\Omega \setdef {\bigcup_{ijk} \omega_{ijk}}$, where
each $\omega_{ijk}$ is a voxel. The greyscale images are further
segmented using the Markov random field segmentation algorithm
\cite{Schlueter14}. The algorithm is able to distinguish the region
occupied by the glass beads as well as that by biofilm. In particular,
each voxel $\omega_{ijk}$ is assigned to one of the domains
$\Omega_r,\Omega_b$, or $\Omega_f$, with a categorical variable
(marker) $r_{ijk}$ as follows
\ba
\label{eq:ijk}
r_{ijk}= \left\{ 
\begin{array}{l}
0: \omega_{ijk} \in \Omega_f \; \mathrm{(fluid)}\\
1: \omega_{ijk} \in \Omega_r\; \mathrm{(rock\ (glass\ beads))}\\
2: \omega_{ijk} \in \Omega_b\; \mathrm{(biofilm)}
\end{array} \right. .
\ea
The boundaries of $\Omega_r$ and of $\Omega_b$ are aligned with the
boundaries of the voxels, which is a necessary approximation.

\begin{figure}
\includegraphics[width=0.8\textwidth]{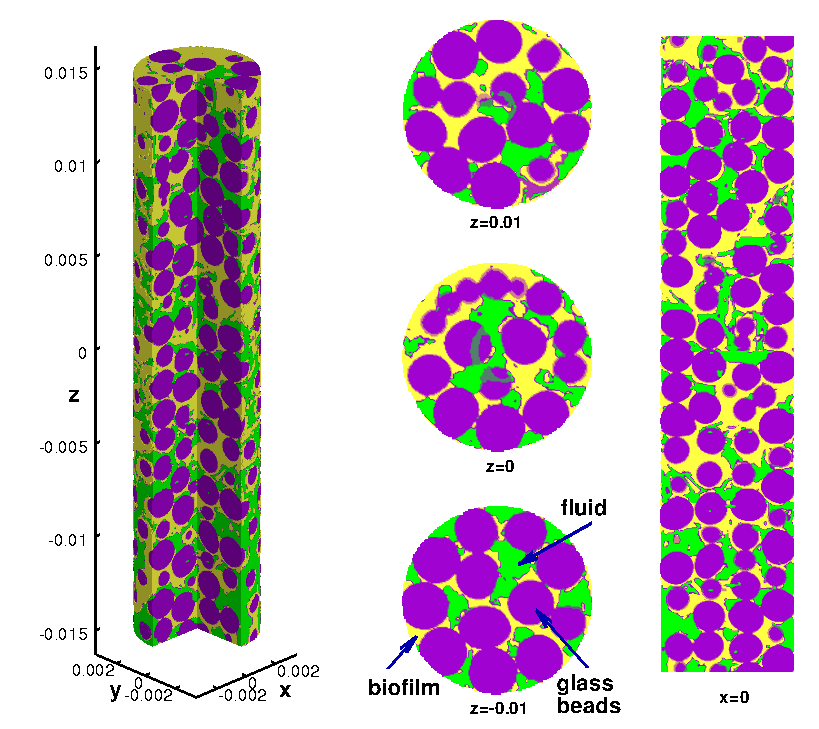}
\caption{Geometry of column $\Omega^{1, red2}$ after two voxel
  reductions. Visible are: the glass beads domain $\Omega_r^{1,red2}$, the
  region $\Omega_b^{1,red2}$ occupied by biofilm, and the region $\Omega^{1,red2}_f$
  occupied by the flowing fluid.
\label{fig-col1-3d-cross}}
\end{figure}

Illustrations in Fig.~\ref{fig-col1-3d-cross} give an idea about the
complexity of the imaging and segmentation process. The voxel grid
shown in Fig.~\ref{fig-col1-3d-cross} is actually a further
approximation discussed next.

\subsection{Grid for flow computations}
\label{sec:grid}

For the needs of computations we need a covering of $\Omega$ by a
computational grid. In particular, the body fitted grids give well
resolved local flow results, but require additional overhead, not
justified by the accuracy of the conductivities; see \cite{PT13}.  A
convenient and practical choice is to use directly the (structured)
grid of voxels $\omega_{ijk}$ covering the cylindrical domain
$\Omega_f$, of size
$
\phi_0^* \times 610 \times 610 \times 2833 \times 0.25 \pi \approx \mathrm{330M}
$. 
Since some passage between the glass beads may be as small as of
single voxel size, we require at least one level of computational grid
refinement by dividing each voxel into 8 computational cells.
However, the grid of $330M \times 8$ cells calls for a significant
computational effort in flow and transport models, and is very
challenging even only for pre- and postprocessing, e.g.,
visualization. Based on our previous experience in \cite{PT13} we
perform therefore some reduction of the original voxel grid by voxel
coarsening and/or by cropping $\Omega$ to a rectangular subdomain
$\tilde{\Omega}$.

Cropping $\Omega$ to a box-shaped region $\tilde{\Omega}$ is necessary
for anisotropic conductivity upscaling, and does not significantly
affect the upscaled conductivities as shown later in
Sec.~\ref{sec:darcy} and Tab.~\ref{tab-permxyz}. 

Voxel coarsening is performed by replacing 8 neighboring voxels
$\omega_{ijk}$ with one voxel $\omega^{red 1}_{IJK}$ whose property
$r_{IJK}$ as in \eqref{eq:ijk} equals that of the majority of
$r_{ijk}$ of the aggregated voxels. This corresponds to a rather
simple upscaling with which the computational effort decreases by
about an order of magnitude. In this paper we used two reduction
steps so that the length $h^{red2}_{\omega} = 2 \times 2 h_{\omega}$
of each voxel $\omega^{red 2}_{IJK}$ is $h^{red2}_{\omega}=42\cdot
10^{-6}$~m, where ``red \#'' denotes the level of reduction.

After voxel reduction is complete, we assign the computational grid
cells denoted by $\Omega_{pqr}$ to be either identical to
$\omega^{red2}_{IJK}$, or to their refinement.  In particular, for 3D
flow simulations we use the computational grid $\TT_h$ with
$
h= 0.5 h^{red2}_{\omega},
$
so that 8 computational cells $\Omega_{pqr}$ subdivide one of
$\omega^{red2}_{IJK}$.  In the coupled transient
hydrodynamics-biomass-nutrient (H-BN) simulations we use a
coarse grid over small 2D subdomains of $\Omega$ with
$
h=h^{red2}_{\omega}
$.

There is some concern that the strategies of cropping and reduction
may reduce the quality of the flow computations. In our previous study
in \cite{PT13} on glass-beads, sandstone, and synthetic geometries, we
determined that the voxel reductions, e.g., from $red1$ to $red2$,
yield an increase in the conductivities by 7\% to 37\%.  On the other
hand, grid refinement decreases the conductivities by 4\% to 9\% for
each consecutive refinement level. In one study of a simple synthetic
dataset, six levels of grid refinement led to about 25\% reduction in
conductivity with respect to that computed for the initial grid. In
other words, the increase due to voxel reduction is somewhat mitigated
by the decrease due to the grid refinement. Overall, while these
effects appear significant, the complexity of the computations on the
original voxel geometries for full columns is prohibitive.

\subsection{Evaluating effect of biofilm growth without simulations}
\label{subsec:eff-bio}

\begin{figure}
\begin{center}
\begin{tabular}{llll}
\includegraphics[width=0.25\textwidth]{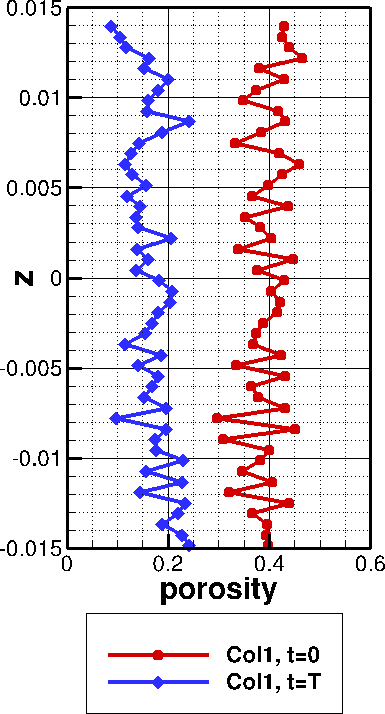}
&
\includegraphics[width=0.25\textwidth]{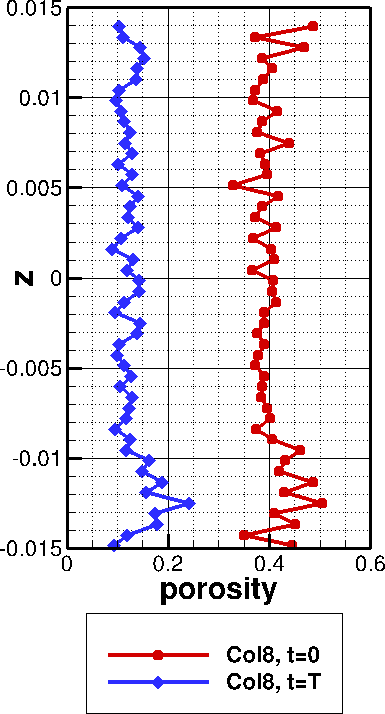}
&
\includegraphics[width=0.25\textwidth]{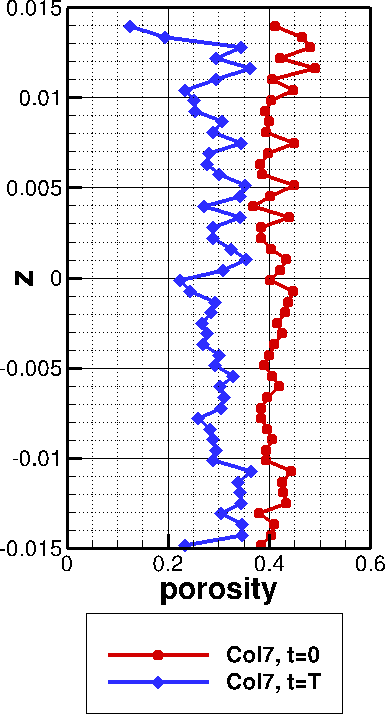}
\\
(a) $\Omega^1$, Re=10&
(b) $\Omega^8$, Re=1&
(c) $\Omega^7$, Re=0.1&
\\
\end{tabular}
\end{center}
\caption{Porosity of fifty slices along the vertical direction ($z$),
  each slice 56 voxels thick, for the original non-voxel reduced geometries,
  at $t=0$ (no biofilm), and at $t=T$ (with biofilm).  The reduction
  in porosity due to biofilm is the largest for high flow rates in
  $\Omega^8$ and $\Omega^1$. The local variations of porosity in
  $\Omega_T$ appear correlated to those in $\Omega_0$.
\label{fig-col-porosity}}
\end{figure}
\myskip{
\begin{figure}
\begin{center}
\begin{tabular}{ccc}
\includegraphics[width=0.2\textwidth]{images/col1_iso_biofilm_narrow_MP_trim2.png}
&
\includegraphics[width=0.2\textwidth]{images/col8_iso_biofilm_narrow_MP_trim2.png}
&
\includegraphics[width=0.2\textwidth]{images/col7_iso_biofilm_narrow_MP_trim2.png}
\\
(a) $\Omega^1$&
(b) $\Omega^8$&
(c) $\Omega^7$
\end{tabular}
\end{center}
\caption{Region $\Omega_f$, i.e., the complement of $\Omega_b$ and
  $\Omega_r$ at $t=T$ for (a) fast (Re=10), (b) intermediate (Re=1),
  and (c) low flow rates (Re=0.1). Note that
  the number of dead-end pores as well as the geometric complexity of
  $\Omega_f$ are increasing with the amount of biofilm, which is
  correlated with high flow rates. See Tab.~\ref{tab-geom} for
  quantitative information corresponding to these illustrations.
\label{fig-col-biofilm}}
\end{figure}
}

\begin{table}
\label{table-columns}
\begin{center}
\begin{tabular}{lcrr|cc|r|r}
\\
\hline\noalign{\smallskip}
&solids&voids&dead-end&$\phi$&$\phi_b$&$ d_{char}$&\# cells\\
&$\abs{\Omega_s}$&$\abs{\Omega_f}$&$\abs{\Omega_d}$&&&&$\abs{\TT_h}$\\
\hline\noalign{\smallskip}
\hline\noalign{\smallskip}
$\Omega_0^1$
& 7338371& 4736720& 615&
0.3922&&1.62&39047712
\\
$\Omega_0^8$
&7222245 & 4863233&398&
0.4024 &&1.58&40538424
\\
$\Omega_0^7$
&7000612&5034880&1590&
0.4182&&1.53&41283216
\\
\hline\noalign{\smallskip}
$\tilde{\Omega}^1_0$
& 4550752&2815280&553&
0.3821 && 1.88&22517816
\\
$\tilde{\Omega}^8_0$
& 4507839& 2858193&498&
0.3879 && 1.83&22861560
\\
$\tilde{\Omega}^7_0$
&4442145&2923887&714&
0.3968&&1.79&23385384
\\
\hline\hline
${\Omega}^1_T$
& 9918224 &2156867&11561&
0.1777&0.183&2.82&17162392
\\
${\Omega}^8_T$
& 10401329 &1684149&16912&
0.1379&0.267&2.97&13256024
\\
${\Omega}^7_T$
&8312777 & 3722715&5682&
0.3088& 0.108&1.83&29736128
\\
\hline
$\tilde{\Omega}^1_T$
&6200908 &1165124 &16912&
0.1559&0.233&3.29&9185696
\\
$\tilde{\Omega}^8_T$
&6428149 &937883 & 51722&
0.1203&0.261&3.30&7503064
\\
$\tilde{\Omega}^7_T$
&5231210 &2134822 &3366&
0.2894&0.109&2.05&17078576
\\
\hline\noalign{\smallskip}
\end{tabular}
\end{center}
\caption{\label{tab-geom}Geometric information derived from data
  sets after two voxel reductions.  The number of cells $\abs{\TT_h}$
  used in flow simulations for cuboid geometries $\tilde{\Omega}^c$
  equals $8(\abs{\tilde{\Omega}^c_f}-\abs{\tilde{\Omega}_d^c})$. For
  $\Omega^c$ the voxel-reduction and cell refinement changes this
  relationship slightly due to the approximation of cylindrical
  boundaries.  The value of $d_{char}$, the ratio of volume of solid
  matrix to the total area of solids [\cite{BearCheng}, p119], is given
  in $10^{-4}$m.  Recall that the column numbers correspond to
  $\Omega^1$ (Re=10), $\Omega^8$ (Re=1), $\Omega^7$ (Re=0.1).}
\end{table}

Some information useful for understanding the biofilm growth and the
relationship between flow rate and the growth can be found without the
flow simulations; see Tab.~\ref{tab-geom}. Here it is useful to
develop additional notation to be refined later. We combine $\Omega_b$
and $\Omega_r$ and call it the ``solid'' domain $\Omega_s$
\ba
\label{eq:omega}
\Omega 
\setdef {\overbrace{\Omega_r \cup \Omega_b}^{\Omega_s} \cup \Omega_f}  
\setdef{\Omega_s \cup \Omega_f},\;\; 
\phi(\Omega) = \frac{\abs{\Omega_{f}}}{\abs{\Omega}},\;
\phi_b(\Omega) = \frac{\abs{\Omega_{b}}}{\abs{\Omega}}.
\ea
The biofilm is excluded from the flow domain, but the biofilm domain
may be involved in some transport processes. Note also that
$\Omega_{s0}=\Omega_r$.

In Tab.~\ref{tab-geom} we confirm that the porosity $\phi_0$
calculated after the voxel reduction step agrees with the experimental
value in \eqref{eq:phi0}; i.e., the imaging and the voxel reduction
preserve the average volume. We also notice that $\phi_0(\Omega)$
exceeds that of $\phi_0(\tilde{\Omega})$ for cropped geometries; this
indicates that there is extra void space near the boundaries of
cylindrical enclosures.

Comparing $\phi_T$ to $\phi_0$ shows the effect of biofilm growth, and
the smallest change is for the slow flow rates. Further insight comes
from studying porosity variations in Fig.~\ref{fig-col-porosity}. We
see, e.g., that the distribution of biofilm is not homogeneous along
columns.

Next, we use a simple algorithm to find the dead-end pores
($\Omega_d$); these are assigned by the imaging to $\Omega_f$, but are
not connected by any path to any of the boundary cells in $\Omega_f$,
and thus are excluded from the flow. In the current model they are
also excluded from transport simulations; we plan to consider
including them in the transport model in the future. A large
number of dead-end pores indicates a more complex structure of a
medium, and in Tab.~\ref{tab-geom} we see an increase in the number of
dead-end pores from $t=0$ to $t=T$, which is most evident for the
large flow rates.

We also calculate $d_{char}$, the characteristic length scale, a
proxy for grain size, which is calculated as the ratio of volume of a
solid matrix to the total area of solids [\cite{BearCheng}, p119]. We
use $d_{char}$ in the Carman-Kozeny correlations discussed later.  The
change in $d_{char}$ from $t=0$ to $t=T$ indicates changes in geometry
which are more pronounced for larger flow rates.

\section{Flow simulations and comparison with experiment}
\label{sec:flow}

Now we discuss the simulations of flow in the columns described in
Sec.~\ref{sec:exp}. The imaging and segmentation provide the geometry
of $\Omega_f\vert_{t=T}$, $\Omega_b\vert_{t=T}$, and
$\Omega_s\vert_{t=T}$ for each column.  By ``subtracting'' out
$\Omega_b$ from $\Omega_s$, we obtain the glass-beads domain
$\Omega_r\vert_{t=0}$ as well as the flow domain
$\Omega_f\vert_{t=0}$. We can thus perform the flow simulations in
$\Omega_{f0}$ and $\Omega_{fT}$ which show the changes in the flow
field due to biofilm growth.  While we cannot compare the simulated
flow field directly to any experiments, we can compare the upscaled
conductivities to those obtained in the physical experiment. We
provide background for these analyses below.

\subsection{Porescale flow model and upscaling}
\label{sec:flowmodel}

Consider a fixed open bounded domain of flow $\Omega_f$,
surrounded partially by a solid region $\Omega_s$.
The flow of liquid (water with nutrients and planktonic cells) in
$\Omega_f$ is assumed to be viscous, and to obey the steady laminar
Navier-Stokes system in $\Omega_f$ for velocity $v$ and
pressure $p$,
\begin{subequations}
\label{eq:ns}
\ba
\rho v \cdot \nabla v - \mu \nabla^2 v &=& - \nabla p, \; x \in \Omega_f,
\\
\nabla \cdot v &=&0, \; x \in \Omega_f.
\ea
We assume no volume forces and ignore gravity. Here $\mu$ is the
viscosity and $\rho$ is the density of the fluid. Other flow models
which may include that in $\Omega_b$ are mentioned in
Sec.~\ref{sec:fluid} but are not implemented here.

The model is complemented by the boundary conditions on $\partial \Omega_f$
\ba
\label{eq:wall}
v \vert_{\Gamma_w \cup \Gamma_0}&=& 0,
\\
\label{eq:inflow}
v \vert_{\Gamma_{in}}&=& v_{in},
\\
\label{eq:outflow}
p\vert_{\Gamma_{out}}&=&0.
\ea
\end{subequations}
The wall condition \eqref{eq:wall} is imposed on internal boundaries
$\Gamma_w \eqdef \partial \Omega_f \cap \partial \Omega_s$. The
external boundary $\partial \Omega_f \cap \partial \Omega$ is divided
into the inflow part $\Gamma_{in}$, the wall no-flow part $\Gamma_0$,
and the outflow part $\Gamma_{out}$, in such a way that $\Gamma_{in}$
and $\Gamma_{out}$ are assigned to a pair of opposite faces of the box
enclosing $\Omega$.  The inlet velocity $v_{in}$ in \eqref{eq:inflow}
is a given constant, and we use the pressure outlet boundary condition
\eqref{eq:outflow}.

To solve \eqref{eq:ns} numerically, we use the ANSYS software
\cite{fluent}, with Finite Volume discretization with grid $\TT_h$
covering $\Omega_f$.
With a fixed $\mu,\rho$, and $\Omega_f$, the only remaining control
parameter is $v_{in}$ and the assignment of the inlet and outlet
boundaries.

\begin{figure}
\begin{center}
\begin{tabular}{ccc}
\includegraphics[height=0.55\textwidth]{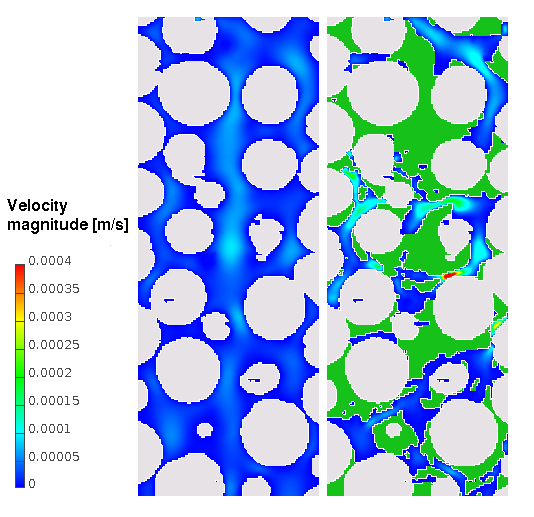}
&
\includegraphics[height=0.55\textwidth]{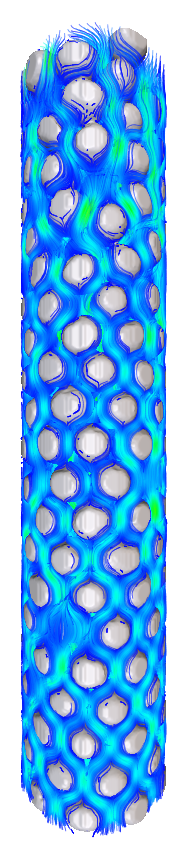}
&
\includegraphics[height=0.55\textwidth]{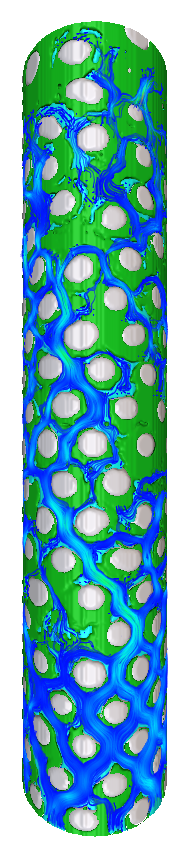}
\\
\hfill a)\hfill b) & c) & d)\\
\end{tabular}
\end{center}
\caption{Visualization of flow simulations in $\Omega^1$. Shown are a)
  the magnitude of velocities in a subsection of $\Omega_{f0}^1$
  (without biofilm) and b) in $\Omega_{fT}^1$ (with biofilm), and c) the streamlines
  of flow in $\Omega_{f0}^1$, and d) the streamlines in
  $\Omega^1_{fT}$. The biofilm is marked in green. The streamlines in
  c)-d) are colored with the magnitude of velocities applied in a)-b)
  and showed in the legend to a). Note the nonhomogeneous distribution
  of biofilm in the vertical direction, and the complexity of flow.
\label{fig-GB}}
\end{figure}

The results of flow simulations are illustrated in Fig.~\ref{fig-GB}
with the contours of velocity magnitude. They are very complex,
especially in geometries with biofilm ($t=T$). In Fig.~\ref{fig-GB} we
see that an increase in the volume of solids due to the appearance of
biofilm domain $\Omega_b$ between $t=0$ and $t=T$ leads to a
substantial reduction of connections among the pores, which, in turn,
influences the directions of the flow.

\subsection{Upscaling flow results}
\label{sec:darcy}

Once the flow simulation is complete, we upscale its results to get
the conductivities $K$ defined for macroscopic pressures $P$ and
velocities $V$ by
\ba
\label{eq:Darcy}
V=K \nabla P= \frac {k}{\mu}\nabla P, \ x \in \Omega,
\ea 
where $K\mathrm{[m^2/Pa \cdot s]}$ is the Darcy conductivity, and $k\mathrm[m^2]$
is the Darcy intrinsic permeability.
For small flow rates \eqref{eq:Darcy} expresses the macroscopic
Darcy's law, and $K$ is a constant. Our upscaling method, developed in
\cite{PTA09} and tested and refined in
\cite{PT10,PTK10,PTS10,TP13,PT13}, calculates $V$ and $\nabla P$ via
volume averaging of $v,p$ over appropriate portions of $\Omega$ and
determines the full tensor $K$ from \eqref{eq:Darcy} applied to
several flow simulations carried out in independent flow directions.
Here we focus on the vertical component of the conductivity $K$
obtained from a single flow simulation in the vertical direction, and
we compare it to that obtained in the experiment from pressure
transducer values. Additional results on anisotropy and varying flow
rates are given in ~\ref{sec:aniso} and \ref{sec:nondarcy}.

\begin{table}
\begin{center}
\begin{tabular}{l|c|c|c|ll}
\hline
&$\phi^{(a)}$&$10^{8} K^*$&$10^{8}K$&$10^{8}K_{CK}$&$10^{8}K_C$
\\
\hline\hline
\multicolumn{6}{c}{Before inoculation, $t=0$, flow rate $500$ml/h}\\
\hline
$\Omega^1_0$
&0.3922&97&186.6 &84.96 &	1020.
\\
$\Omega^8_0$
&0.4024&520&197.2 &90.76 &	1001.
\\
$\Omega^7_0$
&0.4182&70.6&227.2 &100.4&971.9
\\
\hline\hline
\multicolumn{6}{c}{With biofilm, $t=T$, at flow rate $500$ml/h}\\
\hline
$\Omega^1_T$
&0.1777&
&10.42 &13.15 &1408.
\\
$\Omega^8_T$
&0.1379&
&1.246 &6.227 &1216.
\\
$\Omega^7_T$
&0.3088&
&57.84&41.30  &1034.\\
\hline\noalign{\smallskip}
\end{tabular}
\end{center}
\caption{\label{tab-permexp}Conductivity $K\mathrm{[m^2/Pa \cdot s]}$
  in the experiment and from simulations carried out on voxel-reduced
  regions $\Omega^{red 2}$. Shown are the values before inoculation,
  and those computed for geometries $t=T$, at the same flow
  rate. (Experimentally obtained conductivity ratio at the flow rates
  characteristic for each column are given in Tab.~\ref{tab-compare}).
  The right two columns show the Carman-Kozeny $K_{CK}$ and Collins
  $K_{C}$ estimates of $K$ derived from geometrical information in
  Tab.~\ref{tab-geom}. Here $K_{CK}=\frac {k_{CK}} \mu$ and $K_C=\frac
  {k_{C}} \mu$, where $k_{CK}:=0.2 \frac {\phi^3 d^2 _{char}}
  {(1-\phi)^2}$ is given in [\cite{BearCheng}, 4.1.20], and $k_C:=\phi
  d^2 _{char}$ \cite{PT13}. The porosity (a) from Tab.~\ref{tab-geom} 
  shown in the left column is well correlated with the computed values.}
\end{table}

When comparing the simulated and experimental conductivities we face
the following conundrum. Our computations work on fixed geometries and
give the same results when repeated; more broadly, averaging
of computational results over multiple geometries does not appear
natural. On the other hand, the usual experimental practice involves
repeated measurements and reporting the averages as well as the
associated uncertainty.  Should we then compare the averages or rather
the individual column values between experiment and simulation ?  In
this paper we provide and discuss both.

In Tab.~\ref{tab-permexp} we show the values obtained experimentally
before inoculation for each column and those obtained from
simulations for the same flow rate.  Additionally we show the
conductivities obtained by simulations for the geometries with
biofilm, at the same flow rate as that before inoculation. Next, in
Tab.~\ref{tab-compare} we report the ratio of conductivities at $t=T$
to those at $t=0$; the experimental and computational values in
Tab.~\ref{tab-compare} were obtained for the flow rates characteristic
for each column. 

First, we note a large variation in the experimental conductivity
measurements across the columns. In contrast, the simulated
conductivities at $t=0$ were similar to each other across columns, and
they correlate well with the porosities derived from imaging. However,
the $K$ values overpredict the experimentally obtained conductivities
$K^*$ for $\Omega^1$ and $\Omega^7$ and underpredict those for
$\Omega^8$, with the factors ranging from 2 to about 4. These results
appear similar to those in the very recent paper \cite{Scheibe15}
where, depending on the segmentation method and the computational
approach, the discrepancy $\frac{K}{K^*} \approx $ 13.4 to 2.16.  We
recall that in \cite{Scheibe15} the numerical grid corresponds
directly to the voxel geometry, but the sample size made of sandstone
with different size grains was about 10 times bigger in each
direction. The number of cells in the fluid region each direction in
\cite{Scheibe15} was unspecified, but from the information provided we
estimate it to be similar to around 30M as in this paper.

In our paper the discrepancy between $K^*$ and $K$ can be attributed
to several factors.  First, the voxel grid for $\Omega_{f}$ is
obtained by segmentation, with some of its own uncertainty, and the
computational grid is obtained by voxel coarsening and grid
refinement, which introduce further approximations. Second, the flow
model (as any computational model) in $\Omega_f$ gives only
approximate results. Next, the upscaling technique can introduce
additional small discrepancy as discussed in \cite{PTA09} since the
actual REV over which we average is only a subset of the actual flow
region; in this REV we avoid spurious computational pressure values
near the inflow and outflow boundaries.  Additional errors can arise
from an imperfect fitting of the slices/sections during the columns
reconstruction.  Overall, we believe that the agreement between $K$
and $K^*$ is quite good, but there is need for further calibration and
testing.

Interestingly, as concerns averages, the simulated conductivities are,
on average, close to the experimental ones, with a large variation
between the individual column values. The analysis of the data along the
second column in Tab.~\ref{tab-permexp} gives the average conductivity
across the columns $K_0^* \approx 229.2\myskip{ \pm 252}$ to be
close to the simulated conductivities $K_0 \approx
203.7 \myskip{\pm 21.05}$, but we do not expect this closeness to
be a universal phenomenon.

We provide further information to supplement the values of $K$ and
$K^*$.  In Tab.~\ref{tab-permexp} we list the geometrical estimates
$K_{CK}$ and $K_C$ which provide almost consistently the lower and
upper bounds for $K$ and $K^*$. Furthermore, the estimates $K_{CK}$
appear to reflect the changes in geometry due to biofilm clogging
consistently with the porosity changes to biofilm clogging, but the
$K_C$ only provides a stable upper bound. This was already noticed
in \cite{PT13} and indicates the need for further studies towards
reduced models.

Next we discuss the conductivity decrease due to biofilm growth, that
is, we compare $K_T$ and $K_0$.  In Tab.~\ref{tab-compare} we show the
ratio $K_T/K_0$ obtained by numerical simulations and $K_T^*/K_0^*$
estimated from experiment.  The decrease in $K_T$ from $K_0$ is
correlated with a decrease in $\phi_T$ from $\phi_0$; see
Tab.~\ref{tab-geom} and Fig.~\ref{fig-col-porosity}.  The clogging
effects are strong for faster flows (Re=1 and Re=10). The biggest
reduction in conductivity, of about two orders of magnitude, occurs
for $\Omega^8$, i.e., the flow rate Re=1. We hypothesize as in
\cite{Iltis13} that this is due to rapid oxygen consumption at
Re$\geq$1 accompanied by (partial) sloughing in $\Omega^1$ at
Re=10. Similar reduction of conductivity by two or more orders of
magnitude due to clogging, was reported by many authors, e.g,
\cite{Baveye98,Seifert}.

The reduction in $K$ observed in the computations appears similar to
that obtained in the experiment, with the closest agreement for Column
7 (Re=0.1). This reaffirms the need for further calibration and
experiments, but is promising.

\begin{table}
\begin{center}
\begin{tabular}{c|c|c|c}
\hline
&{Re=10}&{Re=1}&{Re=0.1}\\
&${\Omega}^1$&${\Omega}^8$&${\Omega}^7$\\
\hline
$K_T^*/K_0^*$ (experiment)
&0.011&0.092&0.204\\
\hline
$K_T/K_0$ (simulations in 
$\Omega^c$)&0.056&0.063&0.241\\
$K_T/K_0$ (simulations in  $\tilde{\Omega}^c$)& 0.028&0.038&0.205
\\
\hline
\end{tabular}
\end{center}
\caption{\label{tab-compare}Conductivity reduction $K_T^*/K_0^*$ and
  $K_T/K_0$ due to biofilm growth.}
\end{table}

\myskip{
\begin{figure}
\begin{center}
\begin{tabular}{c}
\includegraphics[width=0.8\textwidth]{images/columns_non_Darcy_MP_trim2.png}
\end{tabular}
\end{center}
\caption{Dependence of effective conductivity on flow rate. Results
  are given for datasets $\Omega^1$, $\Omega^8$, and $\Omega^7$ for
  $t=0$ (no biofilm) and $t=T$ (with biofilm).
\label{fig-non-Darcy}}
\end{figure}
}

\section{Biomass-nutrient model coupled to hydrodynamics}
\label{sec:bio}

The growth patterns observed in our experiment as well as in other
biofilm studies require a model for the flow coupled to the biomass
and nutrient advective-diffusive transport, with biomass growth and
nutrient utilization reactions. Biofilm growth occurs by interface
expansion, and there is a certain maximum density of cells in a given
location that cannot be exceeded. Our Biomass-Nutrient model (BN) is
designed to mimick the experiment described in Sec.~\ref{sec:exp},
with the velocity field $v$ computed by the hydrodynamics model (H)
\eqref{eq:ns} described in Sec.~\ref{sec:flow}. The coupled model
(H-BN) aims to reproduce the outcome of the experiments.

The main difficulty is to identify what processes take place in
different parts of the porespace, and how to resolve the free
boundaries between them which change in time. These are overviewed in
Sec.~\ref{sec:fluid}, followed by a literature review in
Sec.~\ref{sec:biolit}. Our mathematical and computational models are
made precise in Sec.~\ref{sec:hbn} and Sec.~\ref{sec:time}.

In this paper we make three simplifying assumptions.  First, we assume
that the biofilm region $\Omega_b$ can only grow, or remain fixed, and
that the change in the flow field is sufficiently slow that solving
(H) and (BN) models via a staggered in time approach is reasonable;
this is justified by the growth patterns observed in the experiment.
Second, we account only for one microbial species {\it (Shevanella
  oneidensis MR-1)} whose mass concentration is denoted by $B(x,t)$,
and for one nutrient only (TSB lumped with DO) whose concentration is
denoted by $N(x,t)$.  Both $B$ and $N$ have units of density. Third,
we assume that the region $\Omega_b$ is impermeable to the flow.  We
model the flow in $\Omega_f$, and the reactive transport in the liquid
region $\Omega_l \setdef{\Omega_f \cup \Omega_b}$ so that
\eqref{eq:omega} is extended as
\ba
\label{eq:omegas}
\Omega \setdef {\Omega_r \cup \Omega_l} 
\setdef{\Omega_r \cup \overbrace{\Omega_b \cup \Omega_f}^{\Omega_l} }
\setdef{\overbrace{\Omega_r \cup \Omega_b}^{\Omega_s} \cup \Omega_f}  
\setdef{\Omega_s \cup \Omega_f},
\ea
reminiscent of the overlapping continua at Darcy scale in
\cite{Ebigbo}.

\subsection{Process description}
\label{sec:fluid}
We describe here the evolution of the domains $\Omega_r, \Omega_f,
\Omega_b$; i.e., the voxel assignment with \eqref{eq:ijk} to one of
these domains can change in time. 

In the experiment the initial
porespace $\Omega_{l0}=\Omega_{f0}$ is inoculated with planktonic
biomass which, as we hypothesize, settle at or close to the walls
$\Gamma_w$ before the pumping of the fluid with the nutrient begins.
At the beginning of the pumping, the flow field obeys \eqref{eq:ns} in
$\Omega_{f}=\Omega_{f0}$. The biomass ``lives'' as planktonic cells
(suspended) in the fluid in $\Omega_{f}$ and is subject to the growth
and (some) advective--diffusive transport in $\Omega_l$.  The
diffusion of biomass in $\Omega_{l}$ is fairly small, since the size of
the cells is large, and the biomass advection is limited, because most
cells adhere to the walls of $\Omega_r$ or to other cells in
$\Omega_b$.  Over time the biomass forms enough of the extracellular
polymeric substance (EPS) to classify the region occupied by the
aggregates of the EPS as $\Omega_b$ which barium cannot penetrate as
in \eqref{eq:barium}; in this region the advective transport
essentially ceases, and diffusion is even further inhibited. The
nutrient is transported in $\Omega_{l0}$ by advection and diffusion,
but its transport in $\Omega_b$ becomes inhibited. As biomass grows,
its amount eventually exceeds the maximum density possible, and the
biomass region has to expand to occupy a larger volume. This occurs by
the interface growth, because only the biomass close to the interface
has access to the bulk of the nutrient transported in $\Omega_f$.

Simultaneously, once the EPS occupies most of $\Omega_b$, the fluid
cannot penetrate $\Omega_b$, and the velocity field needs to be
recomputed. Here we treat the part of $\Omega_b$ filled with the EPS
as impermeable, i.e, part of $\Omega_s$. This impermeability assumption
can be lifted, and there may be {\it some} flow through $\Omega_b$. A
model for such flow may treat $\Omega_b$ as a porous medium or a
region of high viscosity
\cite{Picio_VL_JMemSci_2009,Vrouw_PKL_JMemSci_2010}, but this is
outside the present scope.

The evolution of the fluid domains is realized in (H-BN) as
follows. The time-staggered loop proceeds in steps $t \to t+ \Delta
t$, and is initialized with $\Omega_{l0}=\Omega_{f0} \setdef {\Omega
  \setminus \Omega_{s0}} \setdef{ \Omega \setminus \Omega_r}$.  At
time $t$, given the current solid region $\Omega_s\vert_t$, we find
$v\vert_{t}$ in $\Omega_f\vert_{t}\setdef {\Omega \setminus
  \Omega_s}$, and set $\Omega_l\vert_{t} = \Omega_f\vert_{t}$, i.e.,
we initialize $\Omega_b\vert_t = \emptyset$. Next we solve (BN) in
$\Omega_l$ keeping $v\vert_t$ fixed, and we allow for $\Omega_b$ to
grow. After some $\Delta t$ when the size of $\Omega_b$ increases
relative to that of $\Omega_l$ at the level noticed at the grid
resolution, we pause the simulation. We set $\Omega_s\vert_{t+\Delta
  t} \setdef{ \Omega_s \vert_t \cup \Omega_b\vert_{t+\Delta t}}$,
reset $t+\Delta t \to t$, and start the loop again.
Note that we keep track of the biomass in $\Omega_b$ at all time
steps, even though we do not simulate their evolution once they become
part of $\Omega_s$.

The mathematical models for the growth of $\Omega_b$ are
discussed next.
\subsection{Biofilm models in literature}
\label{sec:biolit}

The biofilm models divide roughly into those at Darcy scale
\cite{Ebigbo}, and those in the bulk fluid
\cite{TangValocchi,coganI,coganII,ZKlapper10,Alpkvist,EE,DE,ED,EPL,EPHL3D,Loos_HEKP_AvanL_2002}.
These models account for the advective--diffusive transport and growth
of biofilm, each in a different, sometimes not fully comprehensive
way. We are not familiar with models which can account simultaneously
for the biofilm dynamics coupled to hydrodynamics in complicated
porescale geometry.

First, most of the models assume that the diffusion coefficient of
biomass $D_B$ is very small or zero, since microbial cells are
typically large, and $D_B$ should be comparable to that of colloids or
large particles.  The models in \cite{EE,DE,ED,Efendiev_EZ_2002} make
a distinction between bulk liquid in which there are no planktonic
cells, and the region in which biomass is nonzero; they further model
the biofilm spreading with diffusivity $D_B(B)$ depending nonlinearly
on $B$, which promotes (infinitely) vigorous spreading of biofilm
close to (some) maximum $B^*$.  This approach of letting $D_B$ blow up
when $B$ is close to some value $B^*$ realizes the maximum constraint,
but appears only heuristic, and makes practical numerical computations
of this singular model very difficult.  On the other hand, in the
discrete models based on cellular automata \cite{EPL,TangValocchi} the
substrate can diffuse everywhere, with $D_B$ reduced by 80\% in
$\Omega_b$. The models calculate the mass in each cell and let the
biomass redistribute so that the total amounts never exceed $B^*$;
however, the results depend on the (heuristic and random) mechanism of
redistribution. Next, the work in \cite{coganI,coganII,ZKlapper10}
includes hydrodynamics and advection, but their general approach of
phase field models with quite complicated nonlinear dependence of
$D_B(B)$ requires detailed resolution at the scale of interfaces
between $\Omega_b$ and $\Omega_f$; these appear unfeasible in porescale
geometries. The approach in \cite{Alpkvist} is to account for dynamic
mass transfer between planktonic and EPS parts of biomass, supported
by a notion of ``pressure'' (equation) which gives an ``advective
velocity'' that drives the interface $\Gamma_{fb}$ between $\Omega_f$
and $\Omega_b$; this approach does not account, however, for the external
velocity field or for transport within $\Omega_b$.  A concept similar
to this ``pressure'' is implicitly implied in the aforementioned
models in \cite{EE,DE,ED}, where the interface $\Gamma_{fb}$ appears
driven by the gradient $\nabla B$, and the nutrient is subject to
transport in $\Omega_l$.

In our model (BN) we allow for biomass and nutrient to be transported
in the liquid phase $\Omega_l$ and for hydrodynamics (H) to be coupled to
(BN) in the time-staggered way outlined in
Sec.~\ref{sec:fluid}. Moreover, we implement the constraint of the
maximum biofilm density in a novel way described next.

\subsection{Biomass evolution and (BN) model}
\label{sec:hbn}

First we recall the well-known growth and consumption rates given by
the Monod expressions
\ba
\label{eq:bmonod}
F(B,N) &=& k_B B \frac{N}{N+N_0}=k_BBg(N),
\\
\label{eq:nmonod}
G(B,N) &=& -k_N B \frac{N}{N+N_0}=-k_NBg(N).
\ea
with $g(N)=\frac{N}{N+N_0}$, where $N_0$ is the Monod constant. The
formula \eqref{eq:bmonod} could be easily extended to account for the
cell death and removal, but this effect is not significant for the
time scale of the experiment discussed here.
The positive constants $k_B,k_N$, and $N_0$ are assumed
known.  

The transport model of $B$ and $N$ is a system of
advecton--diffusion--reaction equations, with diffusion coefficients
$D_B$ and $D_N$, respectively. It has three new elements in contrast to
the literature. 

First, we set up the constraint of maximum biofilm density $B^*$ to
complement these equations
\ba
\label{eq:constraint}
B(x) \leq B^*, x \in \Omega_l
\ea
which is essentially a continuous realization of the biomass spreading
mechanism implemented in the cellular automata models
\cite{EPL,TangValocchi}. (In practice, \eqref{eq:constraint} is
enforced with Lagrange multipliers denoted by $\Lambda$.)

Second, to promote the interface expansion which allows the biomass
growth in spite of \eqref{eq:constraint}, we set the diffusion
coefficient $D_B$ to be dependent on $B$; this is a continuous
realization of the random spreading mechanism implemented in cellular
automata models \cite{EPL}, and is similar to the dependence $D_B(B)$
in \cite{EE,DE,ED}. In our model however we set $\lim_{B\to B^*}
D_B(B)=D^*$, with $D^*$ finite but large, while that in
\cite{EE,DE,ED} was infinite.  In $\Omega_l \setdef{\Omega_f \cup
  \Omega_b}$ the biomass is allowed to diffuse, and we set $\lim_{B
  \to 0} D_B(B)=D_0>0$. (In \cite{EE,DE,ED}, $D_0=0$). In our model
the diffusion term accounts both for the physical molecular diffusion
as well as the mechanism for interface spreading. Our model
is also similar to the concept of internal ``pressure gradient'' in
\cite{Alpkvist} associated with the biomass excess and proportional to
the concentration gradient.

Third, to account for both planktonic cells and EPS, the biomass
density is partitioned between two components
\begin{subequations}
\ba
\label{eq:sum}
B=B^{m}+B^{e}
\ea
where $B^{m}$ and $B^{e}$ correspond to the planktonic mobile biomass
and the EPS, respectively. Their transport is governed by separate
models
\label{eq:Bparts}
\ba
\label{eq:Bm}
\dtt{B^m} +\nabla \cdot(B^m v) 
-\nabla \cdot(D_B\nabla B^m)
\\
\nonumber
+\Lambda =-q + F(B,N),\;  \;  x \in \Omega_l
\\
\label{eq:Be}
\dtt{B^e} -\nabla \cdot(D_B\nabla B^e)=q,\;  \;  x \in \Omega_l.
\ea
Here $\Lambda$ is the Lagrange multiplier needed to enforce
\eqref{eq:constraint}. 

Thus $B^m$ in \eqref{eq:Bparts} is allowed to advect with $v$, whereas
$B^e$ is only allowed to spread and grow from the planktonic cells with
rate $q$. It remains to specify how $B^m$ depends on $B^e$, and this
can be done either via a simple first order rate model or an
equilibrium model
\ba
\label{eq:exchange}
(RATE)\;\;\; q = \nu_0 B^m, {\rm\ or\ }\;\; (EQ)\;\; B^m = \nu_1 B, 
\ea
\end{subequations}
where $\nu_0\geq 0$, and $0 \leq \nu_1 \leq 1$ are constants.
Now the (RATE) model is somewhat similar to (some) principles proposed
in \cite{Alpkvist}. The (EQ) model can be seen as an
approximation of (RATE) if $\nu_0$ is very large. (The case of
moderate $\nu_0$ will be discussed elsewhere).

In summary, by adding \eqref{eq:Bm} to \eqref{eq:Be} and using \eqref{eq:exchange}
(EQ) and with \eqref{eq:sum}, we obtain
\ba
\label{eq:BconB}
\dtt{B} +\nabla \cdot(\nu_1 B v) 
-\nabla \cdot(D_B\nabla B) + \Lambda=F(B,N),\;  x \in \Omega_l,
\\
F(B^*,N)=0, \;\; \nu_1 \vert_{\Omega_b}=0.
\ea
In this model no transport of $B$ (or biomass growth) takes place in
$\Omega_b$ except for the interface spreading.
The interface between $\Omega_{f}$ and $\Omega_{b}$ is defined
implicitly while \eqref{eq:constraint} is enforced; {\bf this} is the
crux of the model. Finally, we can define $\Omega_b$ in a manner
consistent with the intuitive definition \eqref{eq:omegas}
\ba
\label{eq:omegab}
\Omega_b = \{ x: B(x) = B^*\}; \;\; \Omega_f=\{x: B(x) < B^*\}.
\ea
In practice, in the time staggered loop in (H-BN) we set $B_*=\nu_2 B^*$
\ba
\label{eq:omegabp}
\Omega_b = \{ x: B_* \leq B(x) \leq B^*\},\; \; 
\Omega_f \setdef {\Omega_l \setminus \Omega_b},
\ea
with $0 \ll \nu_2 \leq 1$; this classifies more biomass as (mature
EPS) $\Omega_b$ than \eqref{eq:omegab}.

It remains to account for nutrient dynamics  
\ba
\label{eq:BconN}
\dtt{N} +\nabla \cdot(N v) -\nabla \cdot(D_N\nabla N)&=&G(B,N), x \in
\Omega_l.
\ea
To account for highly viscous character of $\Omega_b$, we modify the
diffusivity to depend on $B$, and to be small within and in the vicinity of
$\Omega_b$; see Sec.~\ref{sec:simulations}.

The model (BN) comprises \eqref{eq:BconB} with
\eqref{eq:constraint} and \eqref{eq:BconN}. It requires boundary and
initial conditions for the biomass
\begin{subequations}
\label{eq:bcin}
\ba
\nabla B \cdot n \vert_{\partial \Omega_f \setminus \Gamma_{in} \cup \Gamma_{out}} =0,\;\
B \vert_{\Gamma_{in}} =0,
\\
\label{eq:binit}
B\vert_{\Omega_l,t=t_0}=B_{init}(x),
\ea
as well as for the nutrient
\ba
\nabla N \cdot n \vert_{(\partial \Omega_f \setminus \Gamma_{in}) \cup \Gamma_{out}} =0,\;\
N \vert_{\Gamma_{in}} =N_{in},
\\
N\vert_{\Omega_s,t=t_0}=N_{init}(x).
\ea
\end{subequations}
There may be additional interactions between the biomass and the
boundaries in \eqref{eq:bcin} which cannot be described by the current
no-flux conditions; we plan to consider these in the future. 

\subsection{Numerical model}
\label{sec:time} 

We now describe the numerical discretization of (H-BN) in space and
time. We realize it in a time-staggered fashion, in a sequence of $M$
macro flow time steps $T_{0}<T_1< \ldots T_J < \ldots T_M =T$. In
addition, we set up $N$ micro transport time steps for the (BN) solver,
$t_0<t_1< \ldots t_n \ldots t_N=T$, with the understanding that
$t_0=T_0=0$, and that each flow time step $T_J$ coincides with one of
the transport steps $t_n$.

\subsubsection{Time-discretized coupled (H-BN) model}
We set $T_{0}=0, \Omega_s(T_{0})=\Omega_r$ and
proceed as follows for $J=0,1,\ldots$. 

\begin{enumerate}[STEP I]
\item Given $\Omega_s(T_{J})$, define 
\ba
\Omega_f(T_J)\setdef {\Omega\setminus \Omega_s(T_J)},
\ea
and generate the grid $\TT_h(T_J)$ covering $\Omega_f(T_J)$. (The
resolution $h$ is kept fixed). Set $\Omega_l(T_J) = \Omega_f(T_J)$.
\item Solve the hydrodynamics model (H) \eqref{eq:ns} on $\Omega_f$ to get
  the fluid velocity $v \vert_{T_J}$ using the wall boundary conditions
  \eqref{eq:wall} on $\partial \Omega_f \setminus \Gamma_{in}
  \setminus \Gamma_{out}$ and the inflow and outflow conditions
  \eqref{eq:inflow}, \eqref{eq:outflow}.
\item Project $v \vert_{T_J}$ to a conservative velocity field $v^H
  \vert_{T_J}$ on the grid $\TT_h(T_J)$.
\item Use $v^H \vert_{T_J}$ as well as the initial conditions
  $B\vert_{\Omega_l,T_J}$ as well as $N\vert_{\Omega_l,T_J}$ to solve
  the biomass--nutrient (BN) problem on $\Omega_f$ for $t \in
  (T_J,T_{J+1}]$, where we identify $\Omega_l \setdef{\Omega_f
    \cup \Omega_b}$ as in \eqref{eq:omegabp}. The interface
    $\partial \Omega_l \cap \partial \Omega_b$ is not tracked
    explicitly but can be recovered from the knowledge of $B(x)$.

\item Set 
\ba
\label{eq:omegagrow}
\Omega_s(T_{J+1}) \setdef {\Omega_s(T_J) \cup \Omega_b(T_{J+1})}.
\ea
\item Go to STEP I with $J\rightarrow J+1$. 
\end{enumerate}

Details of the algorithm (STEP I\ldots STEP VI) are as follows. STEP I
is done via simple postprocessing/remeshing. STEP II, as mentioned
before, is done with ANSYS-FLUENT. STEP III is realized using an
algorithm described in \cite{ChippadaDawson}. STEP IV is realized with
the advection-diffusion-reaction model (BN). STEP V requires some
bookkeeping and a restart capability in the (BN) model. 

As concerns the time stepping, our algorithm requires a new macro time
step $T_{J+1}$ to be taken (and a new velocity field $v^H$ to be
recomputed in STEP II), only when the (new) biofilm phase appears,
i.e., $\abs{\Omega_b}>0$, observed at the grid resolution. Thus, while
a macro time step could be, in principle, as small as the (BN) time
step $t_{n+1}$-$t_n$, large macro time steps are taken in practice.

Note that the model accounts in \eqref{eq:omegagrow} only for the
increase of the biofilm phase domain, and is not able to describe a change
$\Omega_b$ that may be due to sloughing.  It also ignores the (BN)
dynamics in the biofilm region(s) $\Omega_b$ after they become a part
of $\Omega_s$.  These features are not a limitation in the present
case.

\subsubsection{Finite volume/CCFD discretization}

The implementation described here is based on the cell-centered finite
differences (CCFD), i.e., the finite volume implementation on
rectangles, with the well known principles established in
\cite{peaceman-book}.  The equivalence of CCFD to the lowest order
Raviart-Thomas spaces on rectangles as well as the associated mass
conservation properties are described in \cite{RT}; see also the
details of implementation in irregularly shaped domains in
\cite{PJW02,PY08}.

We first define the finite volume/CCFD discretization, restricting the
presentation to 2D. In STEP 1 at every macro time step
$(T_M,T_{M+1}]$, the domain $\Omega_l\setdef {\bigcup_{ij}
    \Omega_{ij}}$, where each grid cell $\Omega_{ij}$ is a rectangle
  of center $(x_i,y_j)$ connected to its neighbors, or with one or
  more edges on the boundary $\partial \Omega_l$.

In what follows we consider the vectors $B^n$ and $N^n$ of the
cell-centered unknowns $B_{ij}^n$, and $N_{ij}^n$ which approximate
$B(x_i,y_j,t_n)$ and $N(x_i,y_j,t_n)$, respectively, for each cell
$\Omega_{i,j} \in \TT_h (\Omega_l)$; we also consider the vector
$\Lambda^n$ of the cell-wise Lagrange multipliers.

The fully discrete counterpart of the model \eqref{eq:BconB},
\eqref{eq:BconN} under constraint \eqref{eq:constraint} is, given
$B^n,N^n$, to solve for $B^{n+1},N^{n+1}$ and $\Lambda^{n+1}$, with
$\tau=t_{n+1}-t_n$
\begin{subequations}
\label{eq:snum}
\begin{multline}
\label{eq:Bnum}
\frac{B^{n+1}-B^n}{\tau} +\nabla_h \cdot(B^{n} \nu_1 v) 
+D_B^h(B^{n})B^{n+1} \\+ \Lambda^{n+1} =F(B^{n+1},N^{n}),\;  
\end{multline}
\begin{multline}
\label{eq:Nnum}
\frac{N^{n+1}-N^n}{\tau} +\nabla_h \cdot(N^{n} v)  
+D_N^h(B^{n})N^{n+1}=G(B^{n+1}_{ij},N^{n+1}_{ij}).
\end{multline}
\end{subequations}
This system of nonlinear equations is solved for the vectors
$B^{n+1},N^{n+1},\Lambda^{n+1}$; the additional equation binding
$\Lambda^{n+1}$ to $B^{n+1}$ is explained in
Sec.~\ref{sec:lambda}. Here each of $D_B^h$, and $D_N^h$ is a
positive definite discrete diffusion matrix, which is equivalent, for
constant diffusivity, to the discrete 5-point stencil negative
Laplacian $-\nabla_h^2$. Since in our model diffusivity depends
nonlinearly on the solution, the diffusion matrices do so as well; we
use averaging on the cell edges and time-lagging. The operator
$\nabla_h \cdot$ handles the advective fluxes across the edges by
(first-order) upwinding.

In \eqref{eq:snum} we evaluate the advection term $+\nabla_h
\cdot(B^{n} \nu_1 v)$ explicitly in time, as in the framework of
operator splitting scheme as in \cite{WDos}. This requires
\ba
\label{eq:cfl}
\tau \leq
\tau_{CFL}\eqdef 0.5\min_{ij}\max(h_x/v_x,h_y/v_y)\vert_{\Omega_{ij}},
\ea
where the grid dimensions $h_xh_y = \abs{\Omega_{ij}}$ are as in
Sec.~\ref{sec:grid}, and where the velocities $v_x,v_y$ are the
maximum across the left and right, and bottom to top, edges of
$\Omega_{ij}$, respectively. In \cite{WDos,TRCHEM} the advection is
followed by reaction and next by diffusion. In our scheme we first
handle (explicit) advection for both components, and then combine the
reaction and diffusion steps, while accounting for
\eqref{eq:constraint}, and time-lagging the diffusion coefficients. In
fact, we first solve \eqref{eq:Bnum} for $B^{n+1}$ using the
time-lagged value of $N^n$ in the reaction term, and then solve
\eqref{eq:Nnum} using that new value.

\subsubsection{Implementing the inequality constraint}
\label{sec:lambda}

The main challenge of the biofilm model is the presence of the free
boundary $\Gamma_{fb} = \partial \Omega_f \cap \partial \Omega_b$
between the fluid and biofilm, i.e., the real interface ``seen'' by
the imaging equipment thanks to the barium-based agent. Accounting for
$\Gamma_{fb}$ is essential to describe biofilm growing volumetrically
through the interface. In our model $\Gamma_{fb}$ is defined
implicitly by \eqref{eq:omegab} along with \eqref{eq:constraint}. 

To realize \eqref{eq:constraint} as in obstacle problems and
variational inequalities \cite{ItoKunisch,Ulbrich}, one introduces the
Lagrange multiplier $\Lambda^{n+1}$ which is a vector of dimensions
identical to those of $B$. Then we rewrite \eqref{eq:constraint} as
the nonlinear complementarity constraint, where at every grid point
\bas
\label{eq:lambda}
\forall \Omega_{i,j} \in \TT_h(\Omega_l), \;\; 
\left\{ \begin{array}{l} 
\Lambda^{n+1}_{ij} \geq 0,
\\
B^*-B^{n+1}_{ij} \geq 0,
\\
\Lambda^{n+1}_{ij} (B^*-B^{n+1}_{ij}) =0.
\end{array}
\right.
\eas
This constraint can be efficiently implemented using the ``$\min$''
function and is equivalent to
\ba
\label{eq:min}
\min(\Lambda^{n+1}_{ij},(B^*-B^{n+1}_{ij})) =0, \;\; \forall \Omega_{i,j}  \subset \Omega_l
\ea
In short, the Lagrange multiplier is nonzero in the so-called active
set of gridpoints where $B_{ij}=B^*$, i.e., in the (sub)set of
$\Omega_b$.

The equations \eqref{eq:Bnum}--\eqref{eq:Nnum} with \eqref{eq:min} can
be solved with Newton's method. Since the function ``$\min$'' is not
differentiable everywhere, this version of Newton's method is called
semismooth, and is known to be convergent at almost the optimal rate
\cite{Ulbrich}, provided the Jacobian is never singular.  We can prove
the latter based on some further time-stepping constraints; this along
with other mathematical and numerical analyses is discussed in the
companion paper \cite{PTbio}.

\section{Numerical simulations of (H-BN) model}
\label{sec:simulations}

In this Section we present numerical simulations to test the (H-BN)
model in the conditions similar to the experimental setup of
Sec.~\ref{sec:exp}. The simulations help to gain additional insight
unavailable from the experiment alone.

\begin{figure}
\begin{center}
\includegraphics[width=0.9\textwidth]{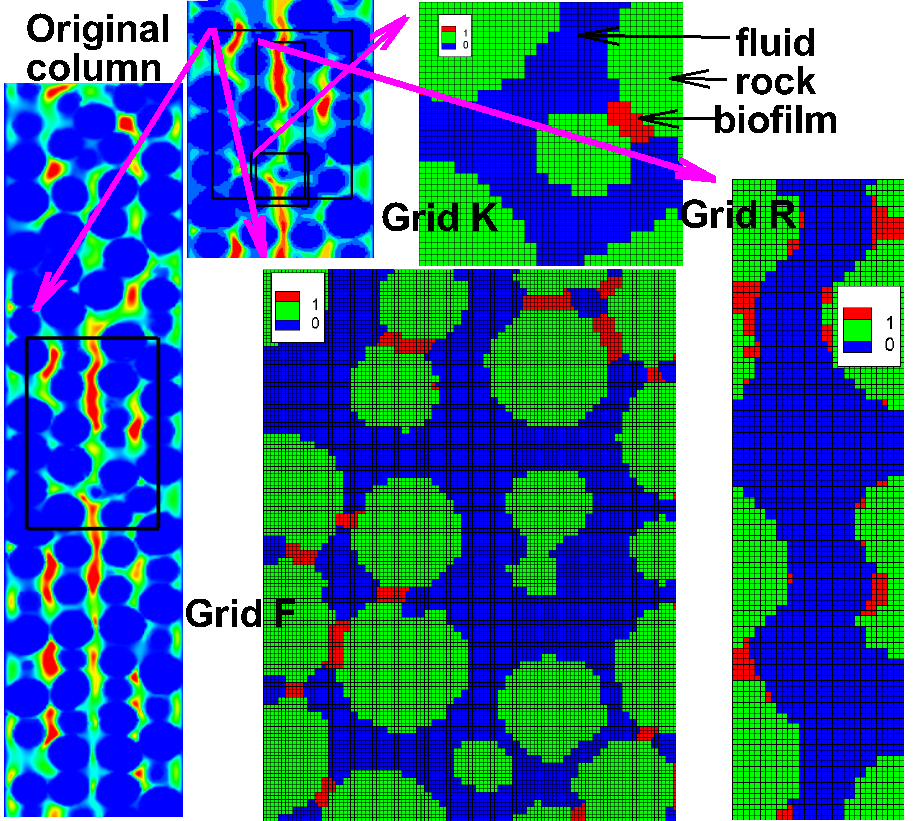}
\end{center}
\caption{Computational 2D domains for (H-BN) and (BN) simulations
  $\Omega^K,\Omega^R,\Omega^F$ are extracted from the cross-sections of
  $\tilde{\Omega}^1$ along the $z$ axis; note the resemblance of
  $\Omega^F$ to the crosssection shown in
  Fig.~\ref{fig-col1-3d-cross}. In $\tilde{\Omega}^1$ shown is the
  velocity magnitude between glass beads. The domains
  $\Omega^K_r,\Omega_r^F,\Omega_r^R$ were slightly conditioned in
  order to assure good percolation. The location of domains
  in $\tilde{\Omega}^1$ is indicated by the arrows connecting the
  upper left corner of each domain. In $\Omega^K$ the arrows indicate
  $\Omega_f$ (blue), $\Omega_r$ (green), and $\Omega_b$ (red). }
\label{fig-bio2fluent}
\end{figure}

Ideally, we would set up time-dependent simulations with enough
resolution to mimick the experimental setup. Unfortunately, this is
not feasible. A quick estimate of the time step required to simulate
(BN) with the flow from hydrodynamics model (H) corresponding to
$v_{in}=10^{-5}$m/s, with grid size of approximately $5\cdot
10^{-6}$m, shows that the time step for explicit advection by
\eqref{eq:cfl} is $\tau_{CFL} \approx O(1)$sec. (Locally the
velocities can be larger than $v_{in}$ which reduces the time step
further). Even with an implicit advection solver, the time steps must
still correspond to the characteristic time for diffusion and
biochemical reactions, and thus are limited. Now, at each time step we
have to solve (at least) two linear systems, each of around $5M$
cells; further difficulties are associated with the nonlinearity
of the problem which requires even more delicate time
stepping. Therefore simulation of $T=11$~days of biofilm growth in the
full column was not possible in this study.

We limit ourselves instead to a focused study aimed to demonstrate the
robustness of the (H-BN) solver, and to the illustration how the
porescale geometries change depending on the model parameters.  We use
only 2D geometries, which are subsets (of a vertical slice) of
$\Omega^{1,red2}$ organized as three cases
$\Omega^K,\Omega^R,\Omega^F$; see Fig.~\ref{fig-bio2fluent}. In
addition, we use $v_{in} \leq 10^{-3}$ because of \eqref{eq:cfl}; this
excludes the fastest flow rate used in the experiment. We also double
the reaction rates given in the literature in order to promote significant
biomass growth within a small $T$ of simulations. Depending on the
case, we simulate only $T=1$~day, or $T=2$~days; these times are usually
sufficient for the clogging to completely block the flow paths.

In addition to the coupled (H-BN) model we also consider the
simulation setup referred to as (BN) in which the velocity values
computed by the model (H) at $t=0$ are fixed, and are not recomputed.

We present the simulation parameters in Tab.~\ref{tab:parameters}.
The top rows include basic parameters common to all the test cases.
Further parameters are listed in bottom rows, along with additional
simulation cases enclosed in brackets, e.g., the case K\_BA uses
geometry $\Omega^K$ and parameters indicated by [BA].

\begin{table}
\begin{tabular}{lll}
\hline
(A)& Growth constant &$k_B=1.8\cdot 10^{-5}$/sec \\
(B)& Utilization constant&$k_N=1.8 \cdot 10^{-4}$/sec\\ 
(C)& Monod constant &$N_0=1.6 \cdot 10^{-3} \mathrm{kg/m^3}$\\
(D)& Nutrient diffusivity& $D_N (x)/D_m= 2\chi_{\Omega_f}(x)+0.1\chi_{\Omega_b}(x)$\\
(E)& Biomass diffusivity& $D_B(x)$, see \eqref{eq:db}\\
(F)& Biofilm phase parameters&$\nu_2=0.9,B^*=0.0012 \mathrm{kg/m^3}$, $B_*=\nu_2B^*$\\ 
(G)& Boundary data&$N_{in} = 0.01 \mathrm{kg/m^3}$, $B_{in}=0$\\
(H)& Initial nutrient&$N_{init}(x) = 0$\\
\hline
(I)& Initial biomass& $\nu_BB_0=0.2$, \ [$\nu_BB_0=0.1$ (SI)]\\
(J)& Inlet velocity &$v_{in}=10^{-5}$m/sec\\
&&[$v_{in}=10^{-4}$ m/sec (FA), $v_{in}=10^{-3}$ m/sec(VFA)]
\\
(K)& Viscous factor &$\nu_1=0$, [$\nu_1=1$ (BA)], \\
&&[$\nu_1=0.5$ (BAP), $\nu_1=0.01$ (BAS)]\\
\hline
\end{tabular}
\caption{\label{tab:parameters}Model and simulation parameters. (A-C)
  adapted from \cite{ZKlapper10,Ebigbo} where we double $k_B,k_N$.
  $D_m=10^{-9}\mathrm{m^2/s}$ is the standard molecular diffusivity in
  water. (D-F) are chosen based on \cite{EE,ZKlapper10}.  (G-H) and
  (I-K) resemble the experimental setup.  (I-K) vary between
  simulation sets as indicated.}
\end{table}

The initial biomass is set to be spread in randomly chosen cells in
the region $\Omega_I$ adjacent to the rock liquid interface $\partial
\Omega_r \cap \partial \Omega_l$. Specifically, enough grid cells
$\Omega_{ij} \subset \Omega_I$ adjacent to $\partial \Omega_r \cap
\partial \Omega_l$ are selected to cover a desired fraction $\nu_{B}$
of $\Omega_I$ with $B_{init}(x) =\nu_B B_0 \cdot 0.0003 \mathrm{kg/m^3}$ for $x
\in \Omega_I$, where $\nu_B B_0$ is a parameter; this choice provides comparable initial
conditions in the different cases $\Omega^R,\Omega^K,\Omega^F$. 

Diffusivity of biomass given in Tab.~\ref{tab:parameters} is given by
\ba
\label{eq:db}
D_B (x)/D_m= .0001\chi_{\Omega_f}(x)+l(B)\chi_{\Omega_b}(x),
\ea
where $l(B)$ is a linear function which increases from $.0001$ at
$B=B_*$ to tenfold $.001$ at $B=B^*$. This choice of $D_B$ is
motivated by \cite{EE,DE,ED} as well as the phase-field models in
\cite{ZKlapper10,coganI}. Note that when $B=B^*$, the diffusion ceases
within most of $\Omega_b$ due to $\nabla B^*=0$, but the interface
growth is promoted to the outside of $\Omega_b$.

\subsection{Simulation results: biomass growth and flow patterns}

\begin{figure}
\begin{center}
\begin{tabular}{ccccc}
&t=10h&t=18h&t=20h&t=22h\\
&\includegraphics[height=0.3\textwidth]{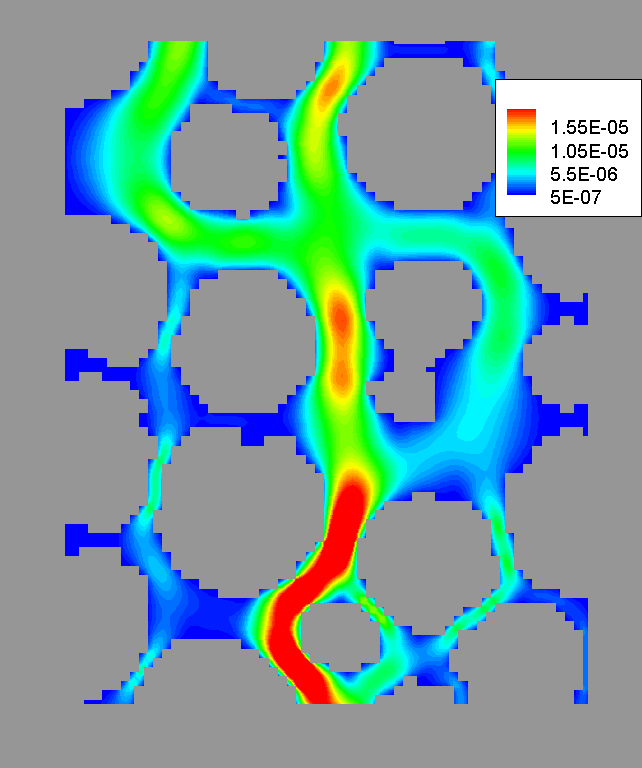}
&
\includegraphics[height=0.3\textwidth]{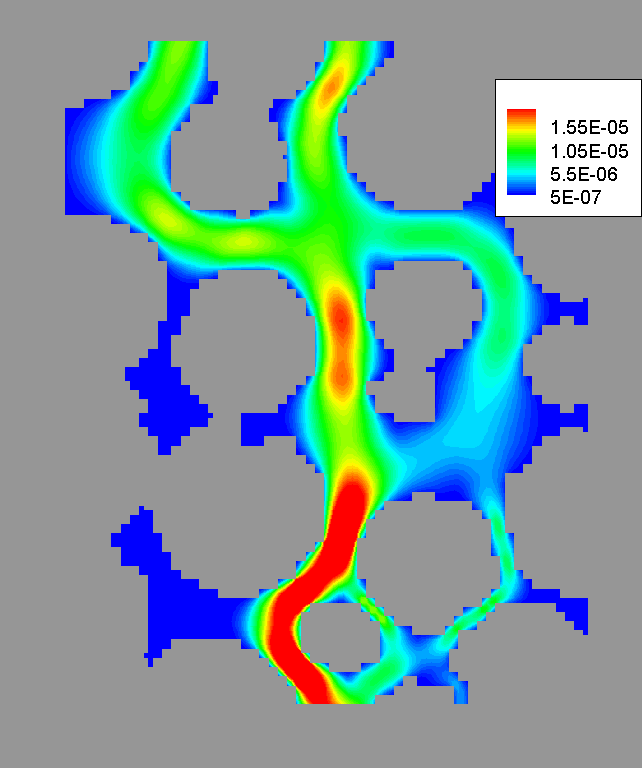}
&
\includegraphics[height=0.3\textwidth]{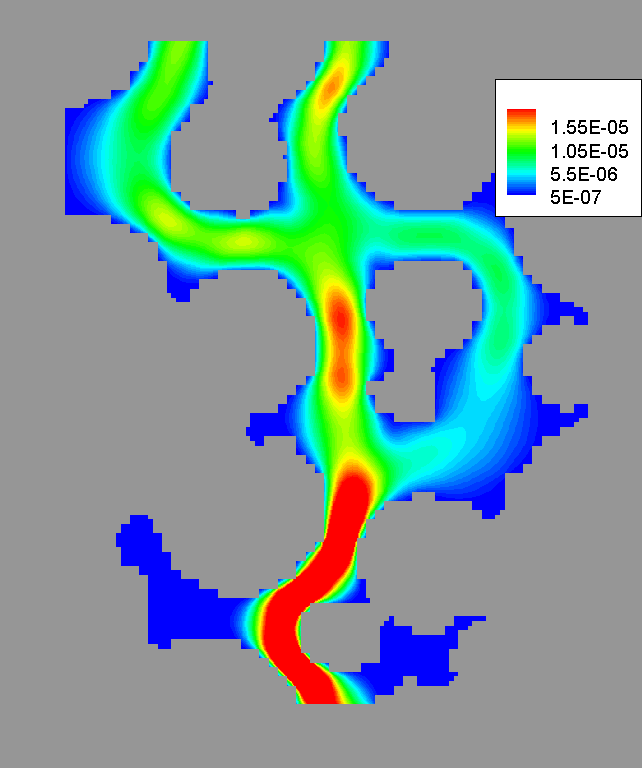}
&
\includegraphics[height=0.3\textwidth]{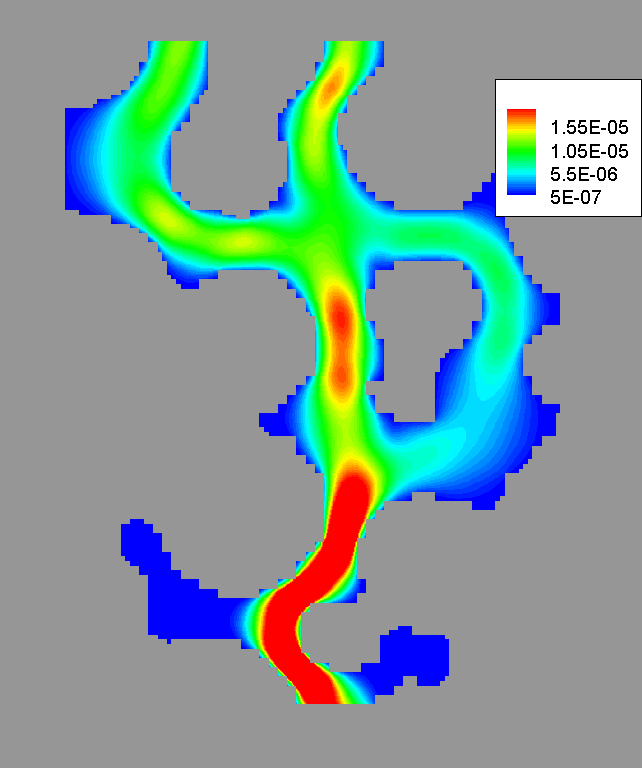}
\\
&\includegraphics[height=0.3\textwidth]{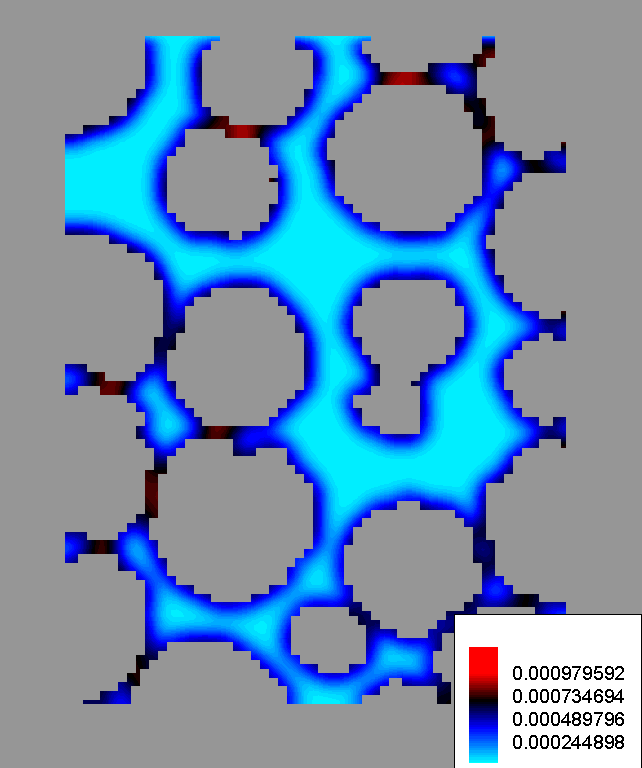}
&
\includegraphics[height=0.3\textwidth]{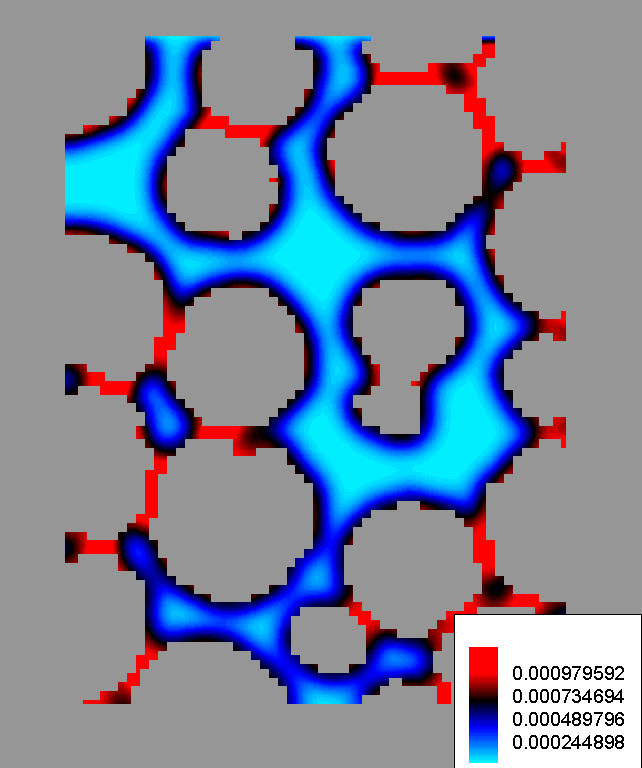}
&
\includegraphics[height=0.3\textwidth]{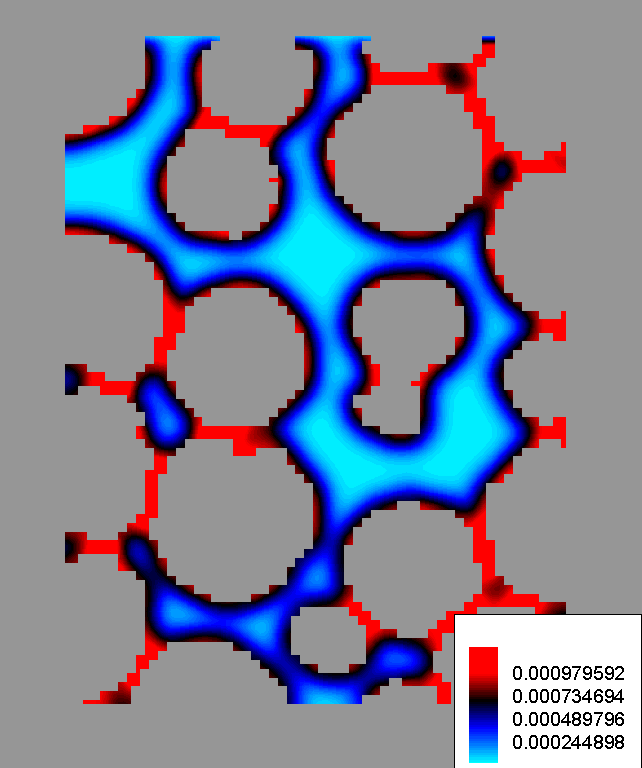}
&
\includegraphics[height=0.3\textwidth]{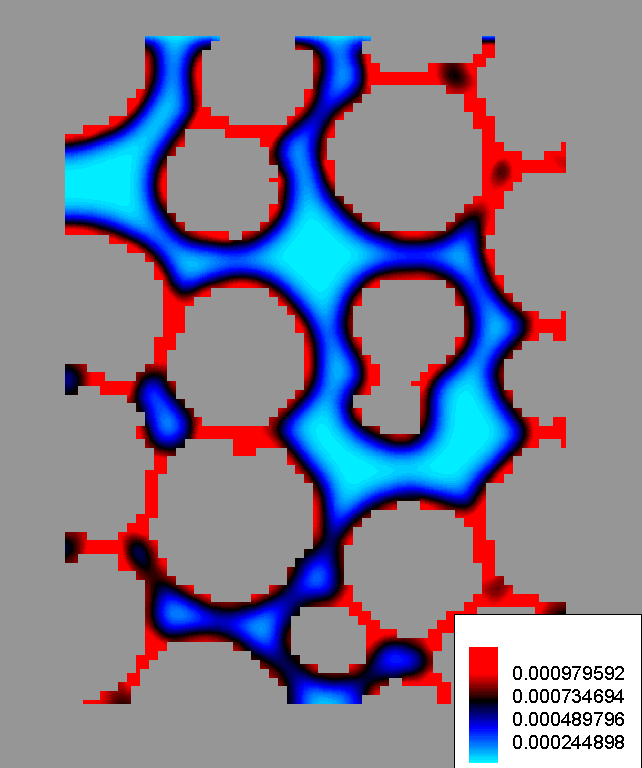}
\\
&\includegraphics[height=0.3\textwidth]{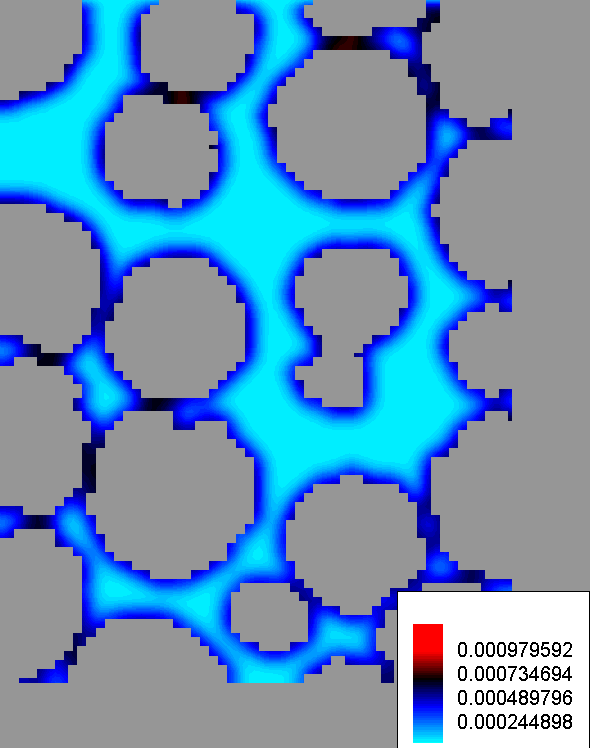}
&
\includegraphics[height=0.3\textwidth]{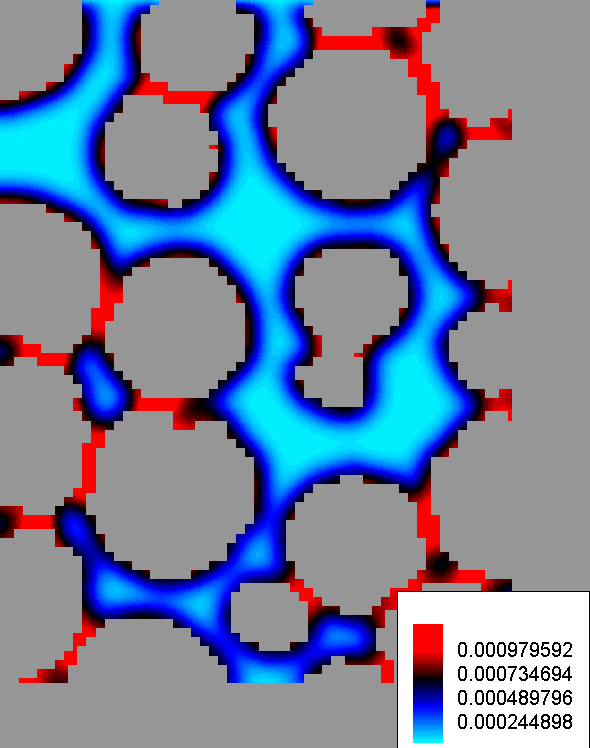}
&
\includegraphics[height=0.3\textwidth]{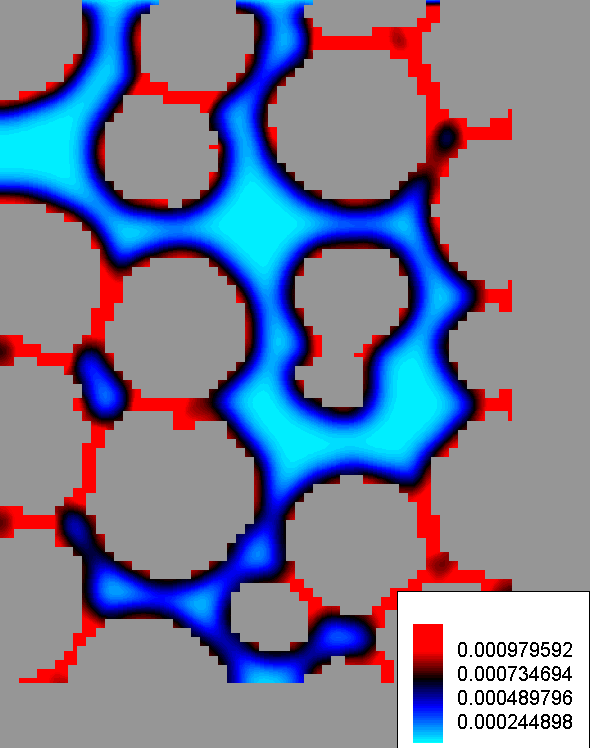}
&
\includegraphics[height=0.3\textwidth]{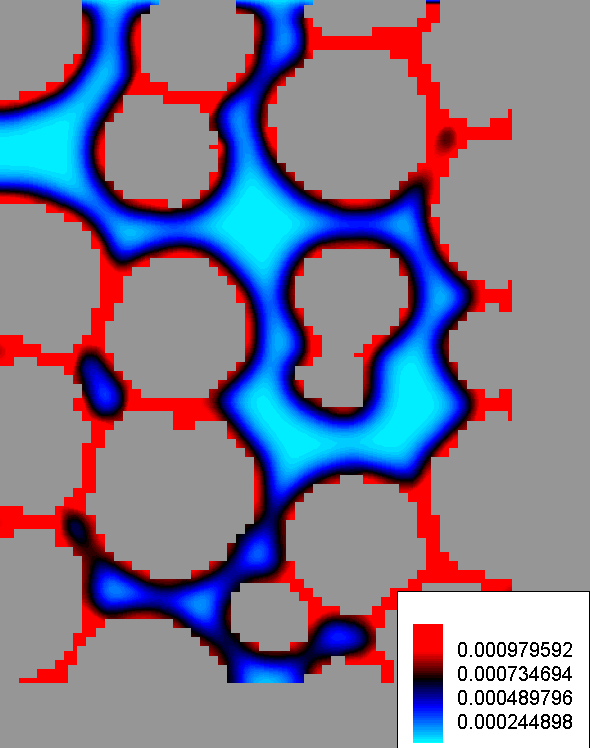}
\end{tabular}
\end{center}
\caption{Velocity computed in (H-BN) simulations (top) show a
  significant change of the flow domain over time. Biomass growth for
  $\Omega^F$ with (H-BN) model (middle) and (BN) model only (bottom)
  show that most of the growth occurs away from main flow
  pathways. However, there is a relatively small difference in the
  distribution of planktonic cells between (BN) and (H-BN) models.
\label{fig:full}}
\end{figure}

\begin{figure}
\begin{center}
\begin{tabular}{cccc}
t=10h&t=18h&t=20h&t=22h\\
\includegraphics[height=0.3\textwidth]{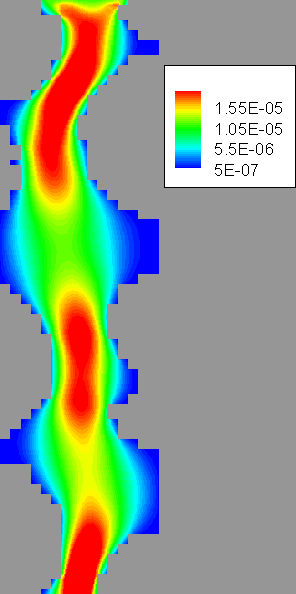}
&
\includegraphics[height=0.3\textwidth]{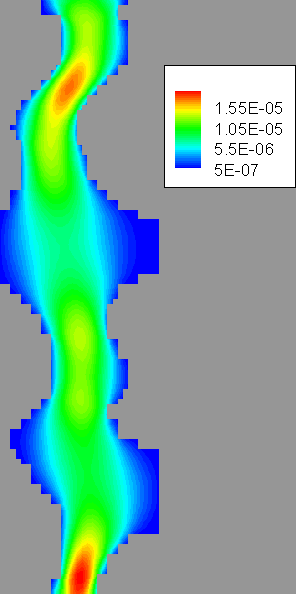}
&
\includegraphics[height=0.3\textwidth]{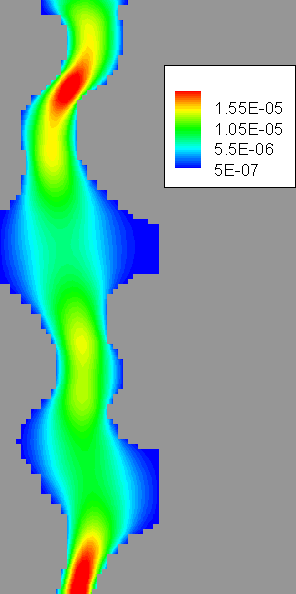}
&
\includegraphics[height=0.3\textwidth]{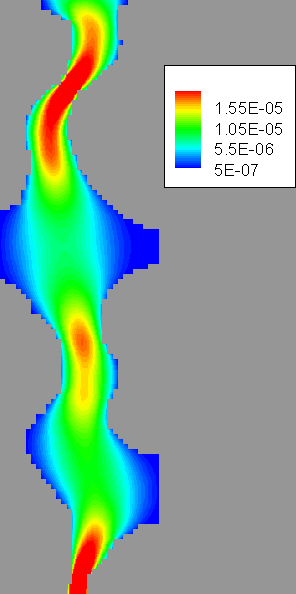}
\\
t=12h&t=18h&t=20h&t=24h\\
\includegraphics[height=0.3\textwidth]{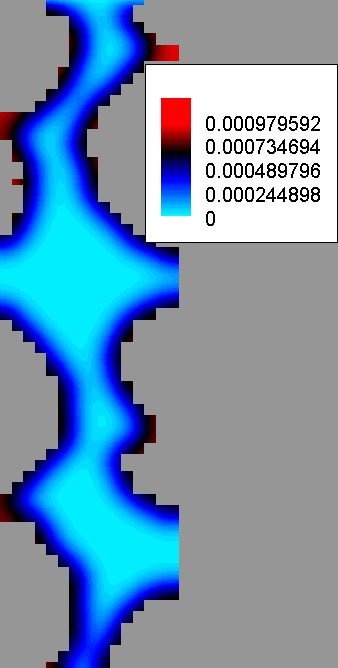}
&
\includegraphics[height=0.3\textwidth]{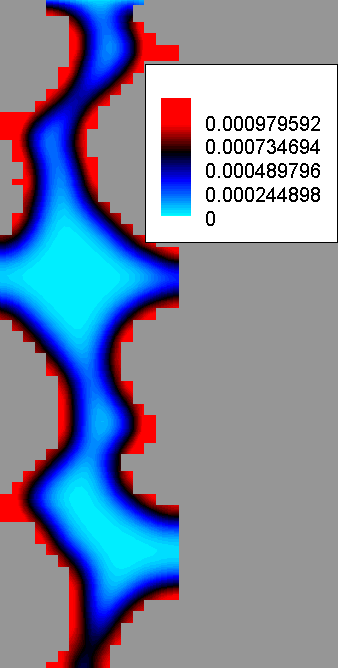}
&
\includegraphics[height=0.3\textwidth]{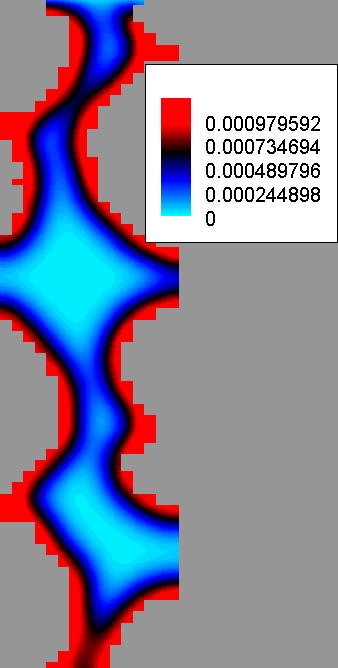}
&
\includegraphics[height=0.3\textwidth]{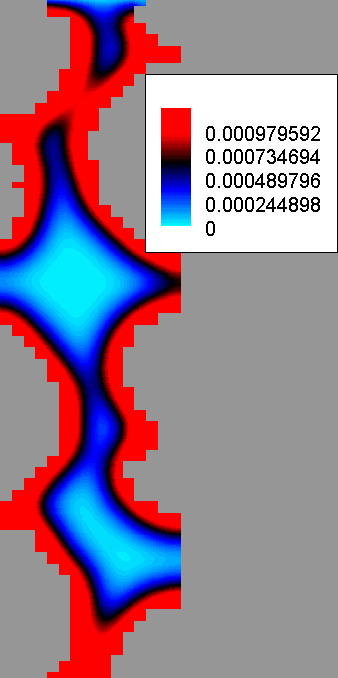}
\\
\multicolumn{4}{c}{$\abs{v}$ (top) and $B$ in (H-BN) simulations}\\
\\
\includegraphics[height=0.3\textwidth]{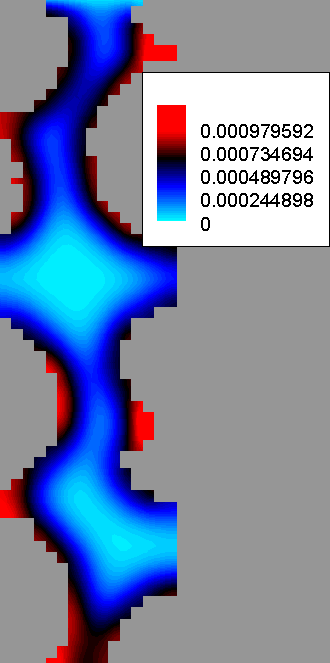}
&
\includegraphics[height=0.3\textwidth]{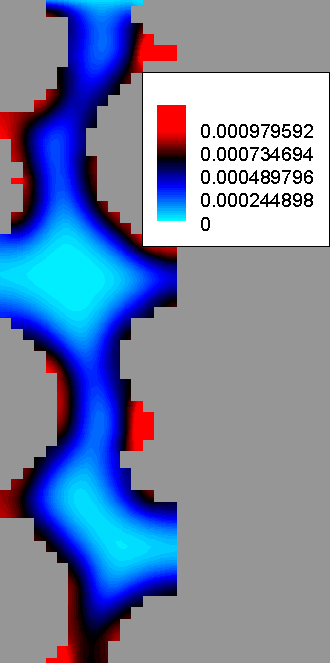}
&
\includegraphics[height=0.3\textwidth]{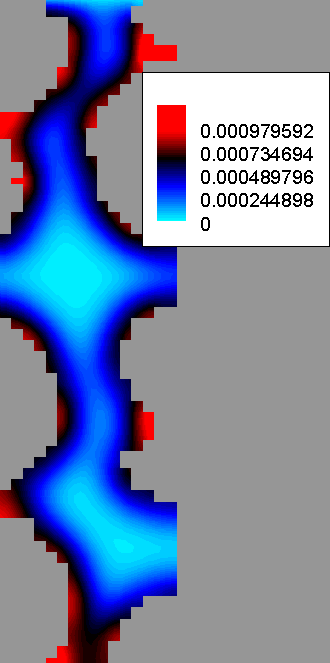}
&
\includegraphics[height=0.3\textwidth]{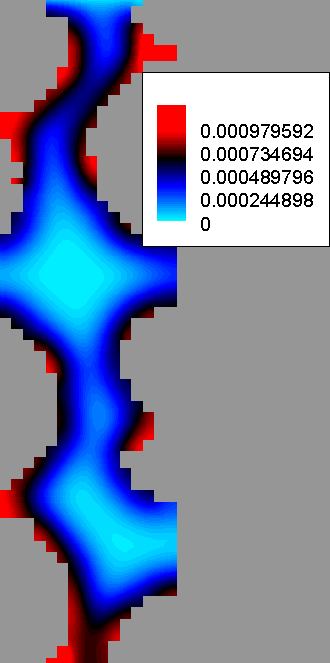}
\\
\multicolumn{4}{c}{$B$ at t=32h simulated from random initial data}
\end{tabular}
\end{center}
\caption{Velocity profiles in $\Omega_f^R$ (top) and biomass amount
  (middle) in (H-BN) simulations for $\Omega^R$; note the domain
  change of $\Omega_f^R$.  Bottom: For smaller initial amount in the
  case [R\_SI], the biomass profiles at $t=32h$ differ only slightly
  between the different random distributions of initial biomass.
\label{fig:rurka}}
\end{figure}

For all the domains $\Omega^F,\Omega^K,\Omega^R$ we set up the flow in
the simulations to be from top to bottom, i.e., the inflow boundary is
the top boundary; plots are shown in Fig.~\ref{fig:full},
\ref{fig:rurka}, \ref{fig:flowrates}. As the simulation progresses, the
biomass grows and the biofilm phase appears at a certain time
indicated in Tab.~\ref{tab:simulations} which varies from 15h to
18h. The regions $\Omega_b$ (and $\Omega_s$) grow as shown in rows
4--5 of Tab.~\ref{tab:simulations}. For some data sets the time 24h
is more than enough for clogging to occur, and the simulation
stops. (The actual experiment discussed in Sec.~\ref{sec:exp}
continues much longer in a larger domain with smaller initial data and
smaller reaction rates.)

\begin{table}
\begin{tabular}{|l|ccc|}
\hline
&$\Omega^K$&$\Omega^R$&$\Omega^F$
\\
\hline
Grid&46$\times$46&32$\times$120&114$\times$152\\
$\abs{\TT_h(\Omega)}$&2116&3840&17328\\
Time $T$&24h&24h&48h\\
\hline
$\abs{\Omega_{s0}}$&1180&1496&9404\\
$\abs{\Omega_{sT}}$, (BN)&1500&3278&13550\\
$\abs{\Omega_{sT}}$, (H-BN)&1553&3491&13975\\
Computational effort$^{(a)}$ (wall clock)&2h&3h&10h\\
\hline
Biofilm phase apppearance (BN)&17h&15h&15h\\
Biofilm phase apppearance (H-BN)&17h&15h&15h\\
Time clogging (BN)&23h&24h&30h\\
Time clogging (H-BN)&28h&24h&29h\\
\hline
\end{tabular}
\caption{\label{tab:simulations}Simulation results for (H-BN) and (BN)
  simulations. (a) Run time for the MATLAB implementation of
  (BN) model}
\end{table}

First we discuss the overall patterns of biomass growth. In
particular, consider Fig.~\ref{fig:full} for the case $\Omega^F$ and
Fig.~\ref{fig:rurka} for $\Omega^R$. With substantial initial amount
$\nu_bB_0=0.2$, we see that the growth appears to take place
preferentially in the more narrow and horizontally aligned passages,
away from the flow in $\Omega^F$, but it occurs essentially uniformly
along the walls in $\Omega^R$. Overall, the patterns in these figures
are qualitatively similar to those obtained in imaging the biofilm
domain as in Fig.~\ref{fig-col1-3d-cross}, but clearly we cannot
hope to match the images from experiment pointwise, e.g., without
knowing the exact distribution of initial biomass.

Now we address the importance of recomputing velocities, i.e., as in
the fully coupled (H-BN) model, where velocities are recomputed each
time the domain $\Omega_f$ changes, compared to the (BN) model, in
which the velocities are computed only once at $t=0$. The velocity
profiles in the changing domain are shown in Fig.~\ref{fig:full},
\ref{fig:rurka}, and the profiles clearly depend on the
domain. However, while the biomass amount shown in
Fig.~\ref{fig:full} has a somewhat different profile between (H-BN)
and (BN) models, this difference is hard to notice at the time scale
and spatial resolution used in our simulations.  The differences are
better seen in Tab.~\ref{tab:simulations} as well as in cumulative
plots in Fig.~\ref{fig:cumulative} discussed below.

Next we assess the effect of a particular initial distribution of
biomass; see Fig.~\ref{fig:rurka}. For a larger initial biomass amount
with $\nu_B B_0=0.2$, the simulated biomass growth (not shown)
corresponding to the different initial conditions appears very
similar. With a smaller amount and $\nu_BB_0=0.1$, the growth shown in
Fig.~\ref{fig:rurka} is more randomly distributed, but the
consistent pattern of biofilm growth away from the flow paths is
preserved; this is likely due to the placement of initial biomass
always next to the walls, even if it is initiated in slightly altered
locations.

\subsection{Effect of flow rates and biomass advection parameter $\nu_1$}

\begin{figure}
\begin{tabular}{ccccc}
\includegraphics[width=0.15\textwidth]{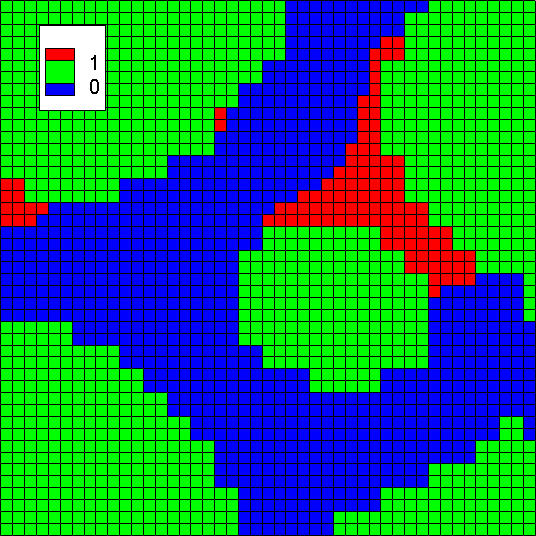}
&
\includegraphics[width=0.15\textwidth]{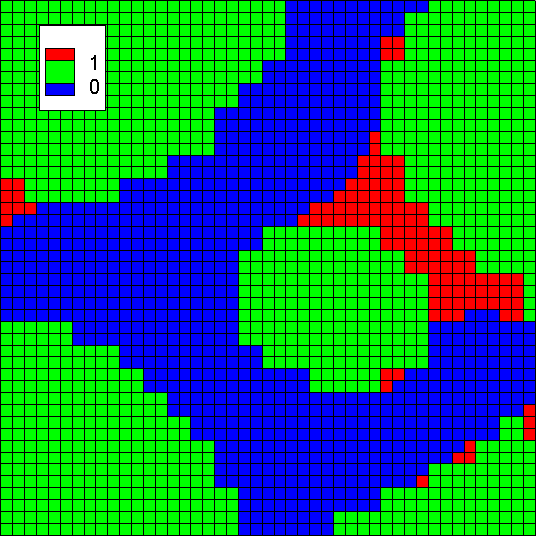}
&
\includegraphics[width=0.15\textwidth]{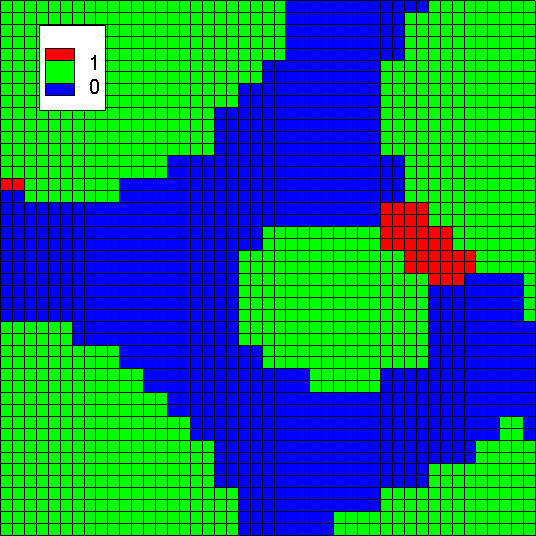}
&
\includegraphics[width=0.15\textwidth]{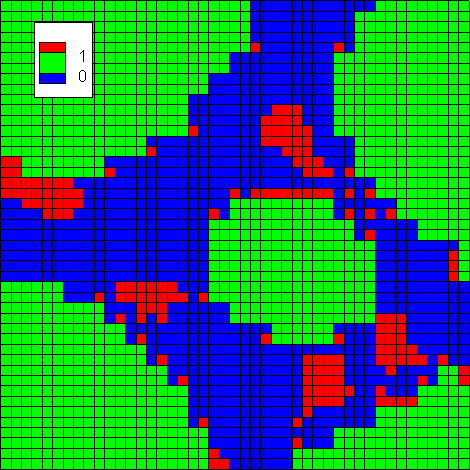}
&
\includegraphics[width=0.15\textwidth]{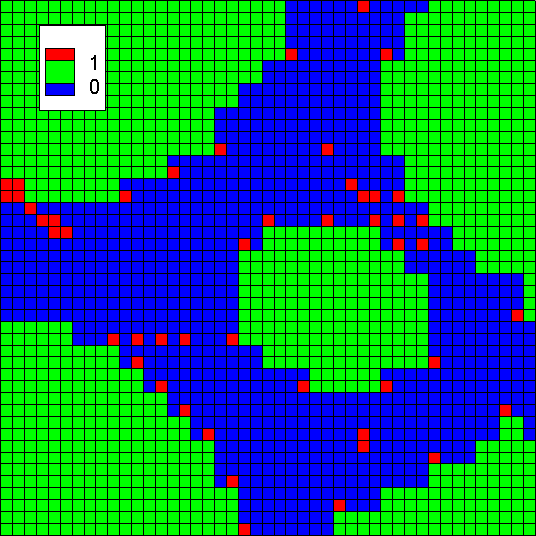}
\\
(a) $\Omega^{K\_VFA}$ &(b) $\Omega^{K\_FA}$&(c) $\Omega^{K}$&
(d) $\Omega^{K\_BAS}$&(e) $\Omega^{K\_BA}$\\
$v_{in}=10^{-3}$&$v_{in}=10^{-4}$&$v_{in}=10^{-5}$&$v_{in}=10^{-5}$&$v_{in}=10^{-5}$\\
$\nu_1=0$&$\nu_1=0$&$\nu_1=0$&$\nu_1=0.05$&$\nu_1=1$\\
\end{tabular}
\caption{The evolution of domain $\Omega^K_f$ with (H-BN) simulations
  depending on the overall flow rate and the advection parameter
  $\nu_1$ in the model \eqref{eq:BconB}. The cases (a-c) show
  faster biofilm growth for faster flow reates which seems to agree
  with the experimental findings. The cases (c-e) show the ability of
  the model to allow the biomass to advect with the flow.
\label{fig:flowrates}}
\end{figure}

Now we discuss the dependence of the growth on the flow rates
and on the modeling assumption concerning the biomass advection. 
In Fig.~\ref{fig:flowrates} (see also later cumulative values in
Fig.~\ref{fig:cumulative}) we compare the base case of
$v_{in}=10^{-5}$ in $\Omega^K$ (see row J in
Tab.~\ref{tab:parameters}) with that when $v_{in}=10^{-4}$ (K\_FA),
or $v_{in}=10^{-3}$ (K\_VFA). The evolution of $\Omega_f^K$ for these
cases indicates that the biofilm phase growth appears concentrated in
the regions of more stagnant flow, and that it increases with higher flow
rates; this seems to agree with the experimental findings.

Second, we simulate cases in which the biomass advection is allowed;
this is controlled by the (ad-hoc) parameter $\nu_1$ from
\eqref{eq:BconB}. We compare the base case where $\nu_1=0$ (no biomass
advection) to the cases with $\nu_1 \neq 0$ (some biomass advection);
the latter cases are listed in row (K) of Tab.~\ref{tab:parameters}
and denoted by (K\_BA), ({K\_BAP}), (K\_BAS), respectively. In the
extreme case of large $\nu_1 \approx 1$ [K\_BA], the biomass behaves
like colloids, and its growth and transport is very different from
that for $\nu_1=0$ (K). The intermediate case (K\_BAS) shows similar
effects which (superficially) resemble sloughing of biomass. 

The experiment suggests that including advection and sloughing is
important, but there is no imaging data to pinpoint exactly the
processes involved, and to help calibrate $\nu_1$. On the other hand,
the models of sloughing available in the literature describe detachment
but cannot yet describe simultaneously the interactions of sloughed
biomass with the walls of the porous medium. While our model results
associated with $\nu_1$ are promising, this aspect needs further work.

\subsection{Cumulative values}

\begin{figure}
\includegraphics[width=0.45\textwidth]{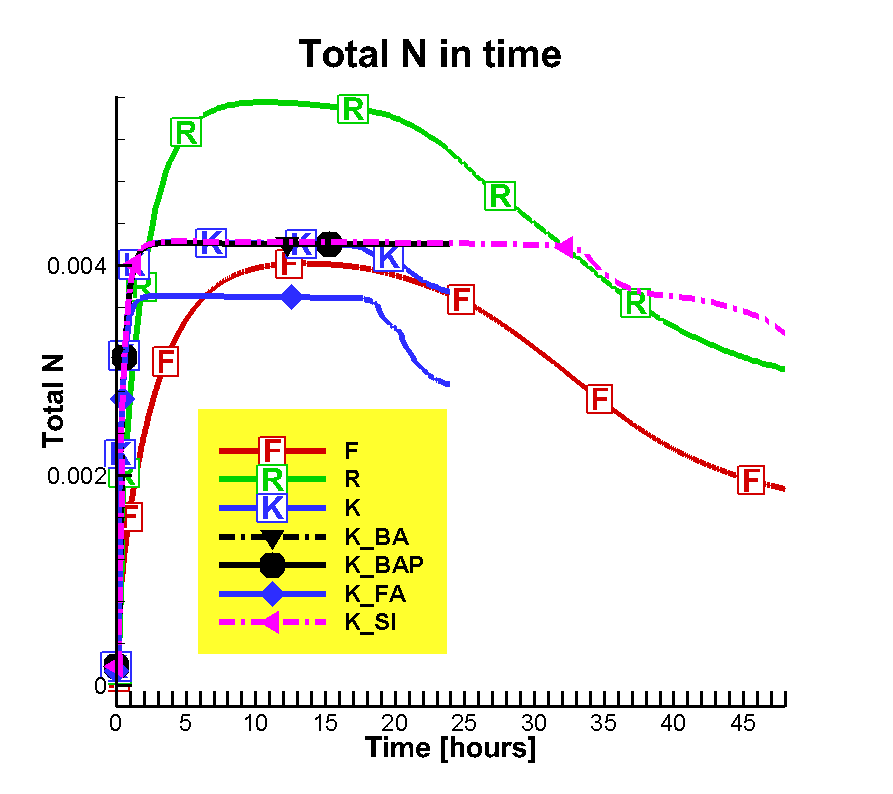}
\includegraphics[width=0.45\textwidth]{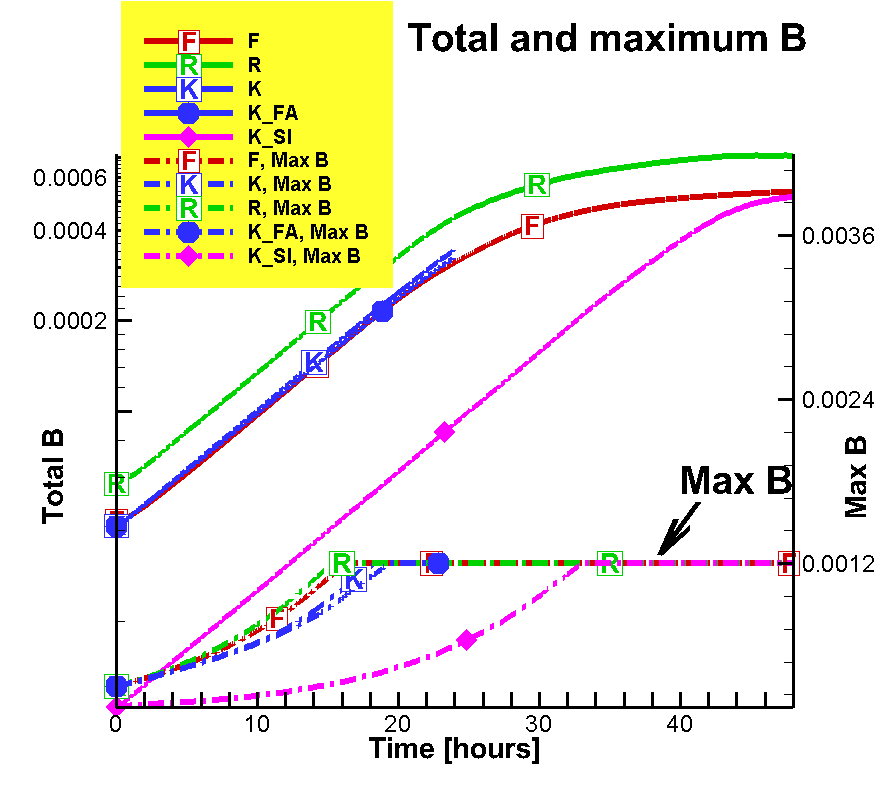}
\\
\includegraphics[width=0.45\textwidth]{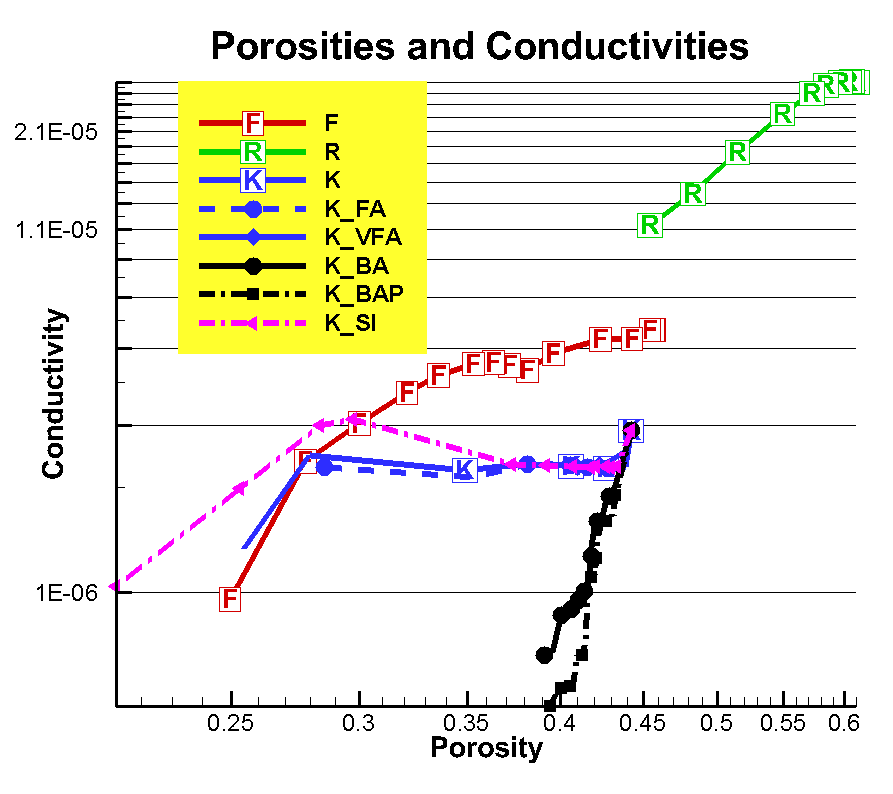}
\includegraphics[width=0.45\textwidth]{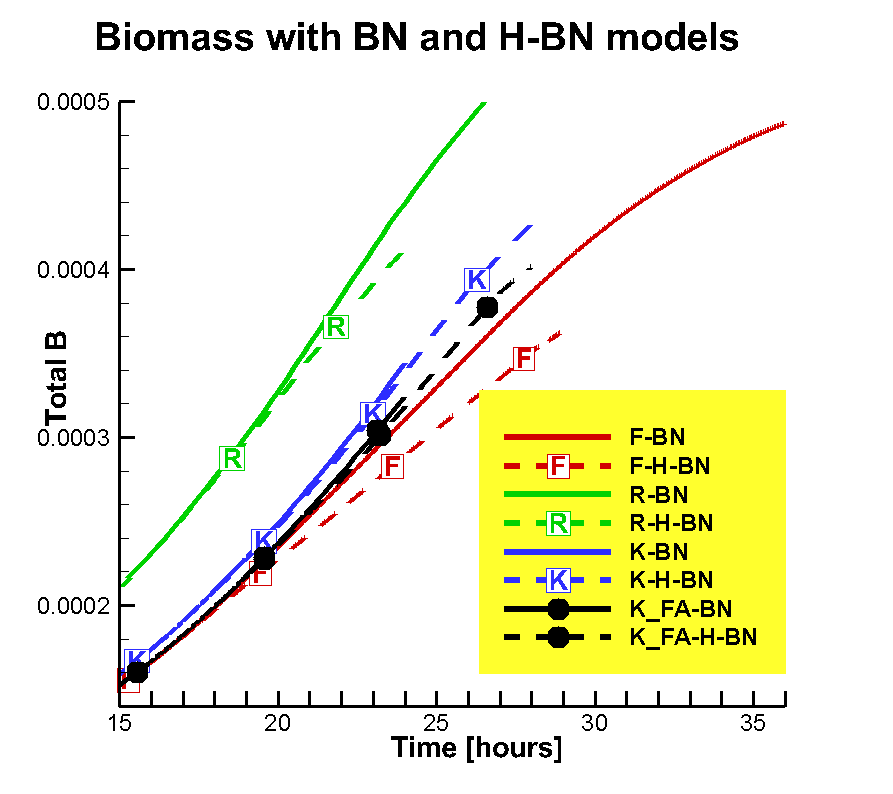}
\caption{Cumulative and average values obtained in the
  simulations. Left top: total nutrient evolution. Left bottom:
  porosity vs conductivities; each marker corresponds to a different
  time snapshot, and some markers are omitted. Right: evolution of
  total and maximum biomass (top) and the zoomed in difference between
  (BN) and (H-BN) models (bottom).
\label{fig:cumulative}}
\end{figure}

The plot of cumulative values of nutrient and biomass dynamics is
shown in Fig.~\ref{fig:cumulative}. First, we notice a very quick
saturation of the domain with the nutrient. This is consistent with
the large (vertical) flow rates combined with the diffusion for
horizontal transport, so that the entire domain appears essentially
filled with nutrient within at most 2h for $v_{in}=10^{-5}$m/s.  The
simulations in the (small) domains $\Omega^F,\Omega^K$, and $\Omega^R$,
suggest that the biomass growth after this initial time is not
nutrient limited. In addition, we notice that about the time the
biofilm phase forms, the nutrient's consumption is substantial, in
spite of being replenished through the advection. In contrast, in the
much larger spatial domains in the experiment, a much longer time is
needed for the nutrient to penetrate the entire domain, and the
optimal growth is for $\Omega^8$ and the intermediate flow rate.

Second, we see in Fig.~\ref{fig:cumulative} that the exponential
growth of biomass tapers off within a few hours after the biofilm
phase appears (when $B^*$ is attained). This is a known phenomenon,
since biofilm tends to grow mostly through interfaces. In our
porescale simulations the growth through interfaces is however more
limited than in the bulk liquid, due to the presence of additional
rock interfaces.

Further, the porosities and upscaled conductivities decrease due to
biofilm growth, but the decrease follows a different pattern for each
of the domains $\Omega^F$, $\Omega^R$, and $\Omega^K$, and for some
model parameters. Since the porosity is a proxy for biomass growth,
the porosity--conductivity plots give the reduced model for dependence
of $K=K(B)$. The case of $\Omega^R$ is the smoothest, and it corresponds
to about cubic decrease in conductivity as a function of porosity;
this is consistent with the ``pipe-like'' character of flow in
$\Omega^R$. For $\Omega^F$ the conductivities decrease in steps
corresponding clearly to the clogging of minor paths.  Finally, the
cases [K\_BA] and [K\_BAP] in which the biomass is allowed to advect
stand out as those corresponding to the most irregular geometry
modifications.  More studies over larger REV are needed to fully
understand these dynamic effects.

\subsection{Further discussion of simulation results}

The simulations and the model are not designed to match the experiment
because of the the tremendous computational effort that would be
required at the time and spatial scales involved. Our results indicate
that the (H-BN) model is robust and gives results which qualitatively
agree with the images obtained in the experiment. However, the
specific growth pattern is strongly dependent on the geometry and on
the initial conditions, and one should not expect a perfect match.

In particular, the experimental results in Fig.~\ref{fig-col-porosity}
show that biomass tends to accumulate generally closer to the inlet
than to the outlet; this can be explained by the relative availability
of nutrient, or by biomass advection \cite{Baveye98}. However, our
simulations (on small domains) do not show that the growth is nutrient
limited but rather that it continues preferentially in narrow
passages. On the other hand, the choice of $\nu_1$ appears important
as it promotes the relocation of biomass towards the outlet.
Finally, our experiments show the dependence of the growth patterns on
the flow rates, even for the limited set of $v_{in}$ chosen, but
further comprehensive parameter studies are needed. Also, one should
perhaps consider some stochastic extensions of the model and
simulations to handle the unknown initial conditions.

\section{Conclusions}
\label{sec:conclusions}

Valuable insight comes from combining imaging, experiments, and
numerical simulations and visualization.  The combined
pore-scale flow experiments and simulations undertaken in this paper
allow the study of the impact of biomass growth on both the micro-scale
flow as well as on the core-scale parameters such as porosity and
conductivity. 

The (H) flow model and upscaling appear robust. We see very good
qualitative agreement and close quantitative agreement between
the experimental and the simulated conductivities. With the simulations, we
obtain detailed understading of flow patterns which is unavailable
experimentally; we also obtain fully anisotropic and heterogeneous
conductivities. Also, simple geometrical correlations provide useful
bounds for conductivities in regular geometries.

The biomass-nutrient (BN) model we proposed and the coupled (H-BN)
model are promising but need more work. Further refinement in close
contact with experimentalists is needed to describe the flow in
$\Omega_b$ as well as the EPS formation and interface growth,
sloughing, the movement of planktonic cells, and the taxis that
attract the cells to interfaces. In particular, the model parameter
$\nu_2$ could be correlated to the estimates of interfacial
area. Next, so far our simulations have shown a mild dependence of the
growth on the flow rates and the associated nutrient (DO) availability
while a stronger one was hypothesized based on the experiment in
\cite{Iltis13,ISWW}; we want to calibrate the model better in order to
understand this fundamental feature. Furthermore, we plan to improve
the efficiency of the (BN) model, e.g., via parallelization, to enable
larger 3D domain studies.
Finally, due to the prohibitive complexity of (H-BN), it is currently
impractical to conduct substantial parameter studies, and we intend to
explore further various reduced models.

While more work is needed, an improved and calibrated (H-BN) model can
be eventually used in a predictive mode to test various scenarios, a
task that cannot be accomplished easily through experimentation of
this complexity.





\section*{Acknowledgments.} 
The authors would like to thank the anonymous referees whose remarks
helped to improve this paper, and in particular for pointing out the
very recent paper \cite{Scheibe15} which describes studies similar
to ours, but for different porous media and at coarser resolution.

In addition, we would like to acknowledge our funding and other
resources. M.~Peszynska and D.~Wildenschild were partially supported
by the grant NSF DMS-1115827 ``Hybrid modeling in porous media''.
A.~Trykozko received funding from the Polish-Norwegian Research
Programme operated by the National Centre for Research and Development
under the Norwegian Financial Mechanism 2009-2014 within Project
Contract No Pol--Nor/209820/14/2013; her research was also supported
in part by PL-Grid Infrastructure.  We thank Dr. Kerstin Kantiem (ICM,
University of Warsaw) for visualizations performed with VisNow tool
\cite{visnow}. M.~Peszynska thanks Oregon State University students
Adriana Mendoza and Jessica Armstrong for useful discussions on the
early versions of the (BN) model.

D.~Wildenschild and G.~Iltis were also supported from the
Environmental Remediation Science Program (DE-FG02-09ER64734) under
the Department of Energy, Office of Biological and Environmental
Research (BER), grant ER64734-1032845-0014978. The work was performed
at GeoSoilEnviro- CARS (Sector 13), Advanced Photon Source (APS),
ANL. This research used resources of the Advanced Photon Source, a
U.S. Department of Energy (DOE) Office of Science User Facility
operated for the DOE Office of Science by Argonne National Laboratory
under Contract No. DE-AC02-06CH11357.  We thank the staff (in
particular Dr. Mark Rivers) and acknowledge the support of
GeoSoilEnviroCARS (Sector 13), which is supported by the National
Science Foundation - Earth Sciences (EAR-1128799), and the Department
of Energy, Geosciences (DE-FG02-94ER14466). S.~Schuleter was
supported by a Feodor Lynen Fellowship from the German Humboldt
Foundation.

For the flow computations and (H) part of the (H-BN) model we used an x86
cluster Hydra, HP BladeSystem/ Actina based on AMD Opteron 2435/Intel
Xeon 5660/AMD Opteron 6132 nodes x86\_64 architecture with 24/32/256
GB of memory, operated at Interdisciplinary Centre for Mathematical
and Computational Modelling, University of Warsaw. The (BN) solver was
implemented in MATLAB as a modification of
flow-advection-diffusion-reaction code and supported by the NSF grants
DMS-1115827 and DMS-0511190.

\appendix
\section{Anisotropic conductivities and at large flow rates}
\label{sec:appendix}

\subsection{Computations of full tensor on cropped rectangular domains}
\label{sec:aniso}
\begin{table}
\begin{center}
\begin{tabular}{l|cll|c|ll}
\hline &$10^{8} K_{xx}$&$10^{8}K_{yy}$&$10^{8}K_{zz}$&
$10^{8}K$&$10^{8}K_{CK}$&$10^{8}K_C$ \\ \hline\hline
\multicolumn{7}{c}{No biofilm, $t=0$}\\ \hline\noalign{\smallskip}
$\tilde{\Omega}^1_0$
&180.8 &186.2 &227.6 &203.4&102.7 &	1342.
\\
$\tilde{\Omega}^8_0$
&186.2 &195.8 &273.1 &213.0 &104.4 &	1300.
\\
$\tilde{\Omega}^7_0$
&200.2 &212.5 &234.9 &235.0 &110.2  &1279.
\\
\hline\hline
\multicolumn{7}{c}{With biofilm, $t=T$}\\
\hline\noalign{\smallskip}
$\tilde{\Omega}^1_T$
&6.845 &	7.310 &7.791 &5.965 &11.49 &	1685.
\\
$\tilde{\Omega}^8_T$
&1.716 &	1.199&0.949 &0.811 &4.882 &	1305.
\\
$\tilde{\Omega}^7_T$
&57.19 &	59.33 &55.37 &48.14 &40.28 &	1215.
\\
\hline\noalign{\smallskip}
\end{tabular}
\end{center}
\caption{\label{tab-permxyz}Conductivities $K\mathrm{[m^2/Pa \cdot
      s]}$ from simulations in cropped voxel reduced regions
  $\tilde{\Omega}^{c,red2}$ at $t=0$, and $t=T$, and
  $v_{in}=10^{-5}$m/s.  The left columns show the diagonal components
  of the anisotropic tensor upscaled from three independent simulations. The
  middle column shows the conductivity $K$ in $z$ direction computed
  from a single simulation. The right two columns show the
  estimates $K_{CK}$ and $K_{C}$ of $K$ derived from geometrical
  information in Tab.~\ref{tab-geom}.}
\end{table}

Our anisotropic nonlinear upscaling method developed in \cite{PTA09}
and tested and refined in \cite{PT10,PTK10,PTS10,TP13,PT13} calculates
$V$ and $\nabla P$ via volume averaging of $v,p$ over appropriate
portions of $\Omega$, and determines $K$ from \eqref{eq:Darcy}. The
calculation of all components of the full tensor $K$ (which needs not
be diagonal) requires three computational experiments in which we vary
the assignment of inflow and outflow boundaries to align roughly with
the $x$, $y$, and $z$ directions, but requires rectangular shape of
the domain.  We supplement the data in Tab.~\ref{tab-permexp}
calculated for the full cylindrical columns $\Omega$ with additional
information in Tab.~\ref{tab-permxyz} on the anisotropic
conductivities calculated for cropped domains $\tilde{\Omega}$.  We
present $K_{xx},K_{yy}, K_{zz}$ only and skip the off-diagonal
components.  We see that $K$ reported in Tab.~\ref{tab-permxyz} has
generally a somewhat larger value than that in Tab.~\ref{tab-permexp}, i.e.,
full columns $\Omega$ have smaller values of $K$ than the subregions
$\tilde{\Omega}$ for $t=0$ but the opposite is true for $t=T$. While
the latter appears naturally correlated to the associated difference
in porosity between $\Omega_T$ and $\tilde{\Omega}_T$, the former
could be explained by the difference in the flow paths between
cylindrical and box shaped regions. Second, while $K$ computed from
$\Omega_{f0}$ is essentially isotropic, that for $\Omega_{fT}$ is not,
but there is no consistent pattern of anisotropy. This indicates a
strong nonuniform increase in resistivity to the flow due to the
biofilm growth.
%

\subsection{Computations with varying flow rates}
\label{sec:nondarcy}

In addition to conductivities reported in Tab.~\ref{tab-permexp},
Tab.~\ref{tab-permxyz} and Tab.~\ref{tab-compare}, we performed
computational experiments for a wide range of flow rates; see
Tab.~\ref{tab-results}. As is well known, the nonlinear effects in the
flow appear typically at flow rates corresponding to around $Re$=$1$,
even though this nondimensional number may have ambiguous definitions
at the porescale. At macroscale, the onset of inertia effects is
manifested by a decrease in the conductivity defined by
\eqref{eq:Darcy} \cite{PTA09,PT13}.  With the data in
Tab.~\ref{tab-results} we confirm the presence of a linear flow regime
below $v_{in}=10^{-3}$~m/s (e.g., $Re$=$1$). The value
$v_{in}=10^{-3}$~m/s marks the onset of inertia effects with a
decrease in the conductivities.  Nonlinear effects become visible for
$v_{in}=10^{-2}$~m/s ($Re$=$10$), and  become even more pronounced at
$v_{in}=10^{-1}$~m/s ($Re$=$100$).

\begin{table}
\label{table-GB-simul}
\begin{center}
\begin{tabular}{l|ll|ll}
\hline\noalign{\smallskip}
Re& $\tilde{\Omega}^1_0$ & $\Omega^1_0$ & $\tilde{\Omega}^1_T$ & $\Omega^1_T$\\
\hline\noalign{\smallskip}
       0.1 &203.4     &199.3    &5.965     &10.76    \\
     1     &202.8     &198.8    &5.945     &10.75    \\  
    10     &185.3     &186.6    &5.327     &10.42    \\  
   100     &85.67     &96.15    &2.142     &6.803    \\
	\hline\noalign{\smallskip}
& $\tilde{\Omega}^8_0$ & $\Omega^8_0$ & $\tilde{\Omega}^8_T$ & $\Omega^8_T$\\
\hline\noalign{\smallskip}
       0.1 &213.0      &215.1   &0.811     &1.348    \\
         1 &212.0      &214.0   &0.802     &1.343    \\  
        10 &194.3      &197.2   &0.644     &1.246    \\  
       100 &91.74      &97.15   &0.175     &0.572    \\
\hline\noalign{\smallskip}
& $\tilde{\Omega}^7_0$ & $\Omega^7_0$ & $\tilde{\Omega}^7_T$ & $\Omega^7_T$\\
\hline\noalign{\smallskip}
       0.1 &235.0    & 243.4   &48.14    &58.57      \\
         1 &234.1    & 242.2   &48.10    &58.53      \\  
        10 &217.5    & 227.2   &46.43    &57.84      \\  
       100 &107.7    & 117.8   &28.88    &44.91      \\
\hline\noalign{\smallskip}
\end{tabular}
\end{center}
\caption{\label{tab-results} Conductivity $10^{8} K$ at different flow rates.}
\end{table}

{\scriptsize
\begin{table}
\begin{tabular}{ll}
\hline
$\Omega$&Porous domain (cylindrical)\\
$\tilde{\Omega}$&Rectangular (cropped) subset of $\Omega$\\
$\Omega^{k,red1}$&Domain for column $k$ obtained after one voxel reduction\\
$\Omega_f$&Domain of fluid flow\\
$\Omega_r$&Glass-beads domain\\
$\Omega_b$&Biofilm domain\\
$\Omega_s$&Domain excluded from fluid flow\\
$\Omega_l$&Domain where transport takes place\\
\hline
$q_0,q_T$&Values of some quantity $q$ before inoculation, and at the end\\
$q^*,q$&Experimental and computational values of some quantity $q$\\
$\phi,\phi_b$& Volume fraction of domain of flow and of biofilm domain\\
\hline
$k$&Darcy (absolute, intrinsic) permeability $\mathrm[m^2]$\\
$K$&Darcy conductivity $K=\frac{k}{\mu}$ $\mathrm[m^2/Pa \cdot s]$\\ 
$K_{CK}$&Estimates of Darcy conductivity via Carman-Kozeny relationship\\
$K_{C}$&Collins estimates of Darcy conductivity \\
$K_h$&Hydraulic conductivity $K=1.02\cdot 10^{-4}K_h$ \\ 
\hline
\end{tabular}
\caption{Nomenclature in this paper\label{tab:notation}}
\end{table}
}


\bigskip
{\bf REFERENCES}
\bigskip

\bibliography{peszynska,dorthe,biofilm_AT,sdiff_mpesz,mpesz}
\bibliographystyle{plain}




\end{document}